\documentclass[12pt]{article}
\usepackage{epsfig}
\usepackage{amsfonts}
\usepackage{amsmath}
\usepackage{amssymb}
\usepackage{cite}
\begin{document}
\sffamily
\title{Analytic structure and power-series expansion of the Jost function for
the two-dimensional problem}
\author{
S. A. Rakityansky$^{1)}$\footnote{\small e-mail: {\sf rakitsa@up.ac.za}}
\ and
\ N. Elander$^{2)}$\\
$^{1)}${\small\it Dept. of Physics, University of Pretoria, Pretoria 0002,
South Africa}\\
$^{2)}${\parbox{12cm}{\small\it Div.of Molecular Physics,
        Dept. of Physics, Stockholm University,\\
        Stockholm, SE-106 91, Sweden}}
}
\maketitle
\begin{abstract}
\noindent
For a two-dimensional quantum mechanical problem, we obtain a generalized
power-series expansion of the $S$-matrix that can be done near an arbitrary
point on the Riemann surface of the energy, similarly to the standard effective
range expansion. In order to do this, we consider the Jost-function and
analytically factorize its momentum dependence that causes the Jost function to
be a multi-valued function. The remaining single-valued function of the energy
is then expanded in the power-series near an arbitrary point in the complex
energy plane. A systematic and accurate procedure has been developed for
calculating the expansion coefficients. This makes it possible to obtain a
semi-analytic expression for the Jost-function (and therefore for the S-matrix)
near an arbitrary point on the Riemann surface and use it, for example, to
locate the spectral points (bound and resonant states) as the S-matrix poles.
The method is applied to a model simlar to those used in the theory of quantum
dots.
\end{abstract}
\vspace{.5cm}
\noindent
PACS number(s): {03.65.Nk, 03.65.Ge, 24.30.Gd}\\[5mm]
\section{Introduction}
\label{sec.Introduction}
It is probable that an average reader of this journal perceives the one- and
two-dimensional problems as simplified toy models of quantum mechanics. Although
such an attitude has its roots in the standard courses of quantum mechanics,
this perception is far from being adequate. First of all, modern nano-technology
allows us to fabricate the microscopic quantum devices that behave and can be
described as one- or two-dimensional. The two-dimensional tunneling of particles
play an important role in superconductive tunnel junctions and even in some
biological molecules \cite{Krevchik}. Besides that, the corresponding quantum
mechanical problems are not mathematically simple as one may think. Indeed, in
contrast to the motion of a particle in three-dimensional space, the
one-dimensional motion of the same particle on an infinite line is inherently a
two-channel problem, where the channels are the left and right halves of the
line (see, for example, Refs. \cite{Rakitsa2003, Rakitsa2004}). As far as the
two-dimensional scattering is concerned, its amplitude as a function of the
energy has not a square-root but a logarithmic branching point (see Refs.
\cite{Bolle1, Bolle2, Verhaar3, Verhaar1, Verhaar2, Gibson, Klawunn} as well as
the subsequent sections of the present paper). Therefore from both pure
mathematical and practical points of view, the two-dimensional quantum problem
is worthwhile to consider.\\

In the present paper, we focus on solving the two-dimensional problem with the
help of power series that are similar to famous effective-range expansion (where
$\hbar k$  and $\delta_0$ are the collision momentum and the $S$-wave
phase-shift)
\begin{equation}
\label{effr_0}
   k\cot\delta_0(k)=-\frac{1}{a}+\frac12r_0k^2-Pr_0^3k^4+Qr_0^5k^6+\cdots\ ,
\end{equation}
in terms of the so called scattering length $a$, effective radius $r_0$,
etc., introduced long ago in nuclear physics \cite{Bethe}.\\

As we already mentioned, the energy dependent functions of the
two-dimensional problem have a logarithmic branching point at the
threshold. As a result there is a controversy concerning the two-dimensional
analog of Eq. (\ref{effr_0}). Some authors
\cite{Verhaar3, Verhaar1, Verhaar2, Hammer}
define the scattering length $a'$ by including it in the logarithmic term,
\begin{equation}
\label{effr_a_log}
   \cot\delta_0(k)=\frac{2}{\pi}\left(\gamma+\ln\frac{ka'}{2}\right)+
   \frac{r_0^2}{2\pi}k^2+{\cal O}(k^4)\ ,
\end{equation}
(here $\gamma$ is the Euler's constant) while the others \cite{Bolle1, Bolle2}
try to preserve the traditional form of the right hand side of Eq.
(\ref{effr_0}) and move the logarithmic term to the left hand side,
\begin{equation}
\label{effr_a_nonlog}
   \cot\delta_0(k)-\frac{2}{\pi}\left(\gamma+\ln\frac{k}{2}\right)=
   -\frac{2}{a''}+{\cal O}(k^2)\ ,
\end{equation}
where $\ln a'=-\pi/a''$.\\

We look at this problem from a more general point of view. What is actually
done in the original effective range expansion (\ref{effr_0}) is the
constructing of the function $R(E)=k\cot\delta_0(k)$ in which the square-root
branching of $k\sim\pm\sqrt{E}$ at the threshold is compensated
by exactly the same branching of $\delta_0(k)\sim\pm\sqrt{E}$. As a result the
function $R(E)$ does not have branching points and is a single-valued analytic
function of the energy $E\sim k^2$, and therefore can be expanded in a
convergent series $R(E)=a_0+a_1E+a_2E^2+\cdots$, which is given by Eq.
(\ref{effr_0}).\\

From this reasoning a next
logical step immediately follows: the function $R(E)$ can be expanded in a more
general power series $R(E)=b_0+b_1(E-E_0)+b_2(E-E_0)^2+\cdots$ around an
arbitrary complex energy $E_0$ within the domain of its analyticity. In Refs.
\cite{my2009, my2011}, we realized this idea for the three-dimensional
single-channel and multi-channel problems. In doing this, instead of using
$R(E)$, we expanded the analytic single-valued parts of the Jost functions (or
Jost matrices in the multi-channel case) after explicit separation of the
factors that are responsible for the branching.\\

In the present paper, we use the same approach as in Refs. \cite{my2009,
my2011} to obtain similar expansions of the Jost functions for the
two-dimensional problem. First, we analyze the analytic structure of the Jost
functions and split them in the single-valued and logarithmically branching
parts. Then, we derive a set of differential equations that determine the
single-valued parts. And finally, we look for the solutions of these equations
in the form of power series of the energy. The series (\ref{effr_a_log}) and
(\ref{effr_a_nonlog}) together with simple recipes for calculating any number of
their expansion coefficients, are easily obtained from our more general
expansions that are done around an arbitrary complex point $E_0$. Using {\bf
two-dimensional model potential related to quantum dot theory}, we numerically demonstrate the efficiency and
accuracy of the proposed method.

\section{Jost function}
\label{sec.JostFunction}
Radial part $u_\ell$ of the wave function describing the motion of a particle of
mass $\mu$ with the energy $E$ in a circularly symmetric two-dimensional
potential $U(r)=\hbar^2V(r)/2\mu$ obeys the differential equation (a review of
the partial-wave analysis for the two-dimensional scattering is given in the
Appendix \ref{app.decomposition})
\begin{equation}
\label{radialeq}
   \left[
   \frac{d^2}{dr^2}+k^2-\frac{\lambda(\lambda+1)}{r^2}-V(r)\right]
   u_\ell(E,r)=0\ ,
\end{equation}
where $\ell=\lambda+1/2$ is the angular momentum and
$\lambda=-1/2,1/2,3/2,\dots$. At large distances, where $V(r)\to0$, the radial
equation simplifies
\begin{equation}
\label{ASSradialeq}
   \left[
   \frac{d^2}{dr^2}+k^2-\frac{\lambda(\lambda+1)}{r^2}\right]
   u_\ell(E,r)\approx0\ ,
   \qquad r\to\infty\ .
\end{equation}
This is the Riccati-Bessel equation. As its two linearly independent solutions,
we can choose either the Riccat-Bessel and Riccati-Neumann functions
$j_\lambda(kr)$ and $y_\lambda(kr)$, or the two Riccati-Hankel functions
$h_\lambda^{(\pm)}(kr)$. Any other solution of Eq. (\ref{ASSradialeq}) is a
superposition of the two linearly independent solutions. In particular, we can
write the asymptotics of the physical wave function as a linear combination of
the Riccati-Hankel functions,
\begin{equation}
\label{uASS}
   u_\ell(E,r)
   \ \mathop{\longrightarrow}\limits_{r\to\infty}
   \ f_\ell^{(\mathrm{in})}(E)h_{\ell-1/2}^{(-)}(kr)+
     f_\ell^{(\mathrm{out})}(E)h_{\ell-1/2}^{(+)}(kr)\ ,
\end{equation}
where the energy-dependent combination coefficients
$f_\ell^{(\mathrm{in/out})}(E)$ are called the Jost functions. When
$r\to\infty$, the Riccati-Hankel functions represent the incoming and outgoing
circular waves. Indeed,
\begin{equation}
\label{hH}
     h_{\ell-1/2}^{(\pm)}(kr)
     =
     \sqrt{\frac{\pi kr}{2}}
     H_\ell^{(\pm)}(kr)
     \ \mathop{\longrightarrow}\limits_{|z|\to\infty}\ {}
     e^{\pm i(kr-\ell\pi/2-\pi/4)}
     =
     \mp ie^{\pm i(kr-\lambda\pi/2)}\ ,
\end{equation}
where $H_\ell^{(\pm)}(z)$ are the cylindrical Hankel functions. The Jost
functions $f_\ell^{(\mathrm{in/out})}(E)$ are therefore the asymptotic
amplitudes of the incoming and outgoing waves. Since the flux of the particles
is conserving, for real $E$ we have
$|f_\ell^{(\mathrm{in})}(E)|=|f_\ell^{(\mathrm{out})}(E)|$. Actually, these two
functions are related to each other at different complex values of $E$ as well.
Some of such symmetry properties can be established using the semi-analytic
structure of them that is derived in the subsequent sections. It can also be
shown that the partial wave $S$-matrix is the ratio of these functions
\begin{equation}
\label{sff}
    s_\ell(E)=\frac{f_\ell^{(\mathrm{out})}(E)}
    {f_\ell^{(\mathrm{in})}(E)}\ ,
\end{equation}
and that zeros of the Jost function $f_\ell^{(\mathrm{in})}(E)$ are the
discrete spectral points ${\cal E}_n$,
\begin{equation}
\label{spectralEQ}
    f_\ell^{(\mathrm{in})}({\cal E}_n)=0\ ,
\end{equation}
i.e. the bound and resonant states of the system.

\section{Transformation of the radial equation}
\label{sec.transformation}
Our goal is to establish the analytic structure of the Jost function, i.e. to
find such an expression for it where all possible nonanalytic dependencies on
the energy are given explicitly. This can be done if we transform the
second-order radial equation (\ref{radialeq}) into an equivalent set of
first-order equations.\\

The transformation is done using a method which is known in the theory of
differential equations as the variation parameters method
\cite{Brand_diff.eqs, Mathews_diff.eqs}. Following this method, we look for
the unknown function $u_\ell(E,r)$ in the form similar to its asymptotics
(\ref{uASS}), but with the combination coefficients being new unknown functions
of $r$,
\begin{equation}
\label{ansatz}
   u_\ell(E,r)=
     F_\ell^{(\mathrm{in})}(E,r)h_{\ell-1/2}^{(-)}(kr)+
     F_\ell^{(\mathrm{out})}(E,r)h_{\ell-1/2}^{(+)}(kr)\ ,
\end{equation}
where $F_\ell^{(\mathrm{in/out})}(E,r)$ are the new unknown functions. Since
instead of one unknown function, we introduce two of them, they cannot be
independent of each other. In principle, we can impose any reasonable condition
relating them. Looking at the asymptotics (\ref{uASS}), we see that
\begin{equation}
\label{liminout}
    f_\ell^{(\mathrm{in})}(E)=
    \lim_{r\to \infty}F_\ell^{(\mathrm{in})}(E,r)\ ,\qquad
    f_\ell^{(\mathrm{out})}(E)=
    \lim_{r\to \infty}F_\ell^{(\mathrm{out})}(E,r)\ .
\end{equation}
Therefore at large distances our new functions become constants, and we should
have
\begin{equation}
\label{lagrange}
   \left[\partial_rF_\ell^{(\mathrm{in})}\right]h_\lambda^{(-)}(kr)+
   \left[\partial_rF_\ell^{(\mathrm{out})}\right]h_\lambda^{(+)}(kr)
   =0\ .
\end{equation}
As the additional condition imposed on these functions, we demand that the
relation (\ref{lagrange}) is valid not only at large $r$ but at all distances.
In the variation parameters method this condition is known as the Lagrange
condition. In our case this condition makes $F_\ell^{(\mathrm{in/out})}(E,r)$
to be the Jost functions for the potential which is cut-off at the radius $r$
(in the spirit of the variable-phase approach).\\

Substituting the ansatz (\ref{ansatz}) into the radial equation
(\ref{radialeq}) and using the Lagrange condition (\ref{lagrange}) together
with known Wronskian of the Riccati-Hankel functions,
\begin{equation}
\label{wronskian}
   h_\lambda^{(-)}(kr)\partial_rh_\lambda^{(+)}(kr)-
   h_\lambda^{(+)}(kr)\partial_rh_\lambda^{(-)}(kr)=2ik\ ,
\end{equation}
we obtain a set of two first-order equations which are equivalent to the
original radial equation (\ref{radialeq}),
\begin{eqnarray}
\label{eqFin}
   \partial_rF_\ell^{\mathrm{(in)}}
   &=&
   -\frac{1}{2ik}h_\lambda^{(+)}V\left[F_\ell^{\mathrm{(in)}}h_\lambda^{(-)}+
   F_\ell^{\mathrm{(out)}}h_\lambda^{(+)}\right]\ ,\\[3mm]
\label{eqFout}
   \partial_rF_\ell^{\mathrm{(out)}}
   &=&
   +\frac{1}{2ik}h_\lambda^{(-)}V\left[F_\ell^{\mathrm{(in)}}h_\lambda^{(-)}+
   F_\ell^{\mathrm{(out)}}h_\lambda^{(+)}\right]\ .
\end{eqnarray}
The boundary conditions for these equations follow from the requirement that
the wave function must be regular everywhere. In particular, this means that
$u_\ell(E,0)=0$. It seems that this is not the case because both
$h_\lambda^{(+)}(kr)$ and $h_\lambda^{(-)}(kr)$ that are present in the
expression (\ref{ansatz}), are singular at $r=0$. Their singularities, however,
can cancel each other,
\begin{equation}
\label{hhj}
   h_\lambda^{(+)}(z)+h_\lambda^{(-)}(z)=2j_\lambda(z)\ ,
\end{equation}
if they are superimposed with a same coefficient. This can be achieved if both
$F_\ell^{\mathrm{(in)}}(E,r)$ and $F_\ell^{\mathrm{(out)}}(E,r)$ have the same
value at $r=0$,
$$
   F_\ell^{\mathrm{(in)}}(E,0)=F_\ell^{\mathrm{(out)}}(E,0)\ .
$$
Their common value at $r=0$ determines the overall normalization of the wave
function and therefore can be chosen arbitrarily. To be consistent, we chose it
to be 1/2, which makes $u_\ell(E,r)$ to behave near the origin exactly as the
Riccati-Bessel function and thus the solution with the boundary conditions
\begin{equation}
\label{bcondFpm}
   F_\ell^{\mathrm{(in)}}(E,0)=F_\ell^{\mathrm{(out)}}(E,0)=\frac12
\end{equation}
is what is called the regular solution in the theory of three-dimensional
scattering.\\

For our goal of expressing the non-analytic dependencies of the Jost functions
in an explicit form, it is more convenient to re-write the ansatz
(\ref{ansatz}) in terms of the Riccati-Bessel and Riccati-Neumann functions,
\begin{equation}
\label{ansatzAB}
   u_\ell(E,r)=A_\ell(E,r)j_\lambda(kr)-B_\ell(E,r)y_\lambda(kr)\ ,
\end{equation}
and to obtain the corresponding equations for the unknown functions
$A_\ell(E,r)$ and $B_\ell(E,r)$. Since
\begin{equation}
\label{hpmjy}
   h_\lambda^{(\pm)}(z) = j_\lambda(z)\pm iy_\lambda(z)\ ,
\end{equation}
this is most simply achieved by making the following linear combinations of
Eqs. (\ref{eqFin},~\ref{eqFout})
\begin{eqnarray}
\label{AFinout}
   A_\ell(E,r) &=& F_\ell^{\mathrm{(in)}}(E,r)+
                   F_\ell^{\mathrm{(out)}}(E,r)\ ,\\[3mm]
\label{BFinout}
   B_\ell(E,r) &=& i\left[F_\ell^{\mathrm{(in)}}(E,r)-
                   F_\ell^{\mathrm{(out)}}(E,r)\right]\ .
\end{eqnarray}
This gives
\begin{eqnarray}
\label{eqA}
   \partial_rA_\ell &=&
   -\frac{1}{k}y_\lambda V(A_\ell j_\lambda-
   B_\ell y_\lambda)\ ,\\[3mm]
\label{eqB}
   \partial_rB_\ell &=&
   -\frac{1}{k}j_\lambda V(A_\ell j_\lambda-
   B_\ell y_\lambda)
\end{eqnarray}
with the boundary conditions
\begin{equation}
\label{bcondAB}
   A_\ell(E,0)=1\ ,\qquad B_\ell(E,0)=0\ .
\end{equation}

\section{Complex rotation}
\label{sec.complex_rotation}
Suppose that the potential $V(r)$  is cut off at certain radius $r=R$, then the
right-hand sides of the sets of equations (\ref{eqFin},\ref{eqFout}) or
(\ref{eqA},\ref{eqB}) vanish at $r>R$ and thus the derivatives on the
left-hand sides of these equations become zero, i.e. the functions
$F_\ell^{\mathrm{(in/out)}}$ or $A_\ell$ and $B_\ell$ do not
change beyond this point. Therefore, in the spirit of the variable phase
approach, the functions $F_\ell^{\mathrm{(in/out)}}(E,r)$
are the Jost functions for the potential which is cut off at the
point $r$. Generally speaking, when the potential asymptotically vanishes at
large distances, we should expect the convergence of the
limits (\ref{liminout}).\\

Therefore, the Jost functions can be calculated by numerical integration of the
differential equations (\ref{eqFin},\ref{eqFout}) or (\ref{eqA},\ref{eqB}) from
$r = 0$ up to a sufficiently large radius $R$ where the limits (\ref{liminout})
are reached within a required accuracy. This works perfectly for real values of
the energy $E$. However, when we consider complex energies (for example, in
search for resonances), a technical difficulty arises. This difficulty is caused
by the asymptotic behavior (\ref{hH}) of the Riccati-Hankel functions.\\

\begin{figure}[ht!]
\centerline{\epsfig{file=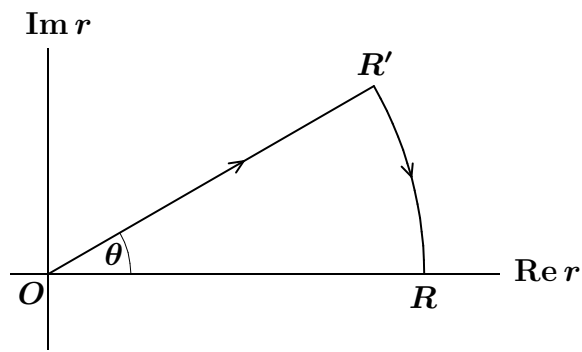}}
\caption{\sf
Deformed contour for integrating differential equations
(\ref{eqFin},\ref{eqFout}) and
(\ref{eqA},\ref{eqB}).
}
\label{fig.pathray}
\end{figure}

When $k$ is complex, either $h_\lambda^{(+)}(kr)$ or $h_\lambda^{(-)}(kr)$
exponentially diverges, depending on the sign of $\mathrm{Im}\,k$. As a result,
either the first or the second of the equations (\ref{eqFin},\ref{eqFout}) does
not give a numerically convergent solution. This difficulty is circumvented by
using the deformed integration path shown in figure \ref{fig.pathray}. Instead
of integrating the differential equations along the real axis from $r = 0$ to $r
= R$, we can reach the final point via the intermediate point $r = R'$ in the
complex plane. Moreover, we can safely ignore the arc $R'R$ since the potential
is practically zero at that distance.\\

Such a complex rotation helps because the asymptotic behavior (divergent or
convergent) of the functions $h_\lambda^{(\pm)}(kr)$ is determined by the sign
of $\mathrm{Im}\,k$. For a given $k=|k|e^{i\phi}$, we can always find such a
rotation angle $\theta$ in $r=|r|e^{-i\theta}$ that the product
$kr=|kr|e^{i(\phi-\theta)}$ has either positive or negative (or even zero)
imaginary part. Various technical details of using complex rotation in
calculating the Jost functions and Jost matrices can be found in
\cite{Rakitsa2003, Rakitsa2004, my2009, my2011, my1996_nuovo, my1997_exact,
my1997_nnn, my1998_LL, my1999_singular, my1999_FBS, my1999_massen, my2007_pade}.

\section{Explicit separation of the non-analytic factors}
\label{sec.factorization}
In order to establish the analytic structure of the Jost functions, we need to
have a closer look at the structure of the Riccati-Bessel and Riccat-Neumann
functions. The following expressions for them (they can be
derived using formulae 9.1.2, 9.1.10, and 9.1.11 of Ref. \cite{Abramowitz}) are
the most useful for this
\begin{eqnarray}
\label{jseries}
   j_\lambda(kr) &=&
   k^{\lambda+1}\sum_{n=0}^\infty k^{2n}f_n^{(\lambda)}(r)\ ,\\[3mm]
\label{yseries}
   y_\lambda(kr) &=&
   k^{-\lambda}\sum_{n=0}^\infty k^{2n}g_n^{(\lambda)}(r)+
   h(k)j_\lambda(kr)\ ,
\end{eqnarray}
where
\begin{equation}
\label{fcoeff}
   f_n^{(\lambda)}(r) =
   \frac{\sqrt{\pi}(-1)^n}{n!\Gamma(n+\lambda+3/2)}
   \left(\frac{r}{2}\right)^{2n+\lambda+1}\ ,
   \qquad\text{for any $\lambda$}\ .
\end{equation}
If $\lambda$ is integer then the expansion of $y_\lambda(kr)$ is also simple:
\begin{eqnarray}
\label{gcoeffinteger}
   g_n^{(\lambda)}(r) &=&
   \frac{\sqrt{\pi}(-1)^{n+\lambda+1}}{n!\Gamma(n-\lambda+1/2)}
   \left(\frac{r}{2}\right)^{2n-\lambda}\ ,\\[3mm]
\label{hzero}
   h(k) &=& 0\ .
\end{eqnarray}
However, for a half-integer $\lambda$, we have a more difficult case:
\begin{eqnarray}
\label{gcoeff}
   g_n^{(\lambda)}(r) &=&
   \left\{
   \begin{array}{lcl}
   \displaystyle
   -\frac{(\lambda-n-1/2)!}{\sqrt{\pi}n!}\left(\frac{r}{2}\right)^{2n-\lambda}
   &,&
   0\leqslant n\leqslant \lambda-\frac12\ ,\\[5mm]
   \displaystyle
   \frac{2}{\pi}\ln\left(\frac{r}{R}\right)f_{n-\lambda-\frac12}^{(\lambda)}(r)
   -&&\\[5mm]
   \displaystyle
   -\frac{(-1)^{n-\lambda-\frac12}\left[\psi(n+1)+\psi(n-\lambda+\frac12)
   \right]}{\sqrt{\pi}n!(n-\lambda-\frac12)!}
   \left(\frac{r}{2}\right)^{2n-\lambda}
   &,&
   \lambda+\frac12\leqslant n<\infty\ ,
   \end{array}\right.\\[5mm]
\label{hnonzero}
   h(k) &=& \frac{2}{\pi}\ln\left(\frac{kR}{2}\right)\ ,
\end{eqnarray}
where $R$ is an arbitrary number (in the units of length). It is arbitrary
because any increase or decrease of $h(k)$ cased by the change of $R$ is
compensated by the corresponding change in the first term of Eq.
(\ref{gcoeff}). The parameter $R$ is introduced to separate the $r$ and $k$
dependencies in the original term containing $\ln(kr/2)=\ln(kR/2)+\ln(r/R)$ and
to have
dimensionless products under the logarithm. In practical calculations the
parameter $R$ can always be taken as the unit of the length, i.e. $R=1$. The
$\psi$-function in Eq. (\ref{gcoeff}) is defined as follows
\cite{Abramowitz}
\begin{equation}
\label{psifunction}
   \psi(n)=\frac{\Gamma'(n)}{\Gamma(n)}=\left\{
   \begin{array}{lcl}
   -\gamma &,& n=1\ ,\\[3mm]
   \displaystyle
   -\gamma+\sum_{m=1}^{n-1}m^{-1} &,& n\geqslant 2\ ,
   \end{array}\right.
\end{equation}
where $\gamma=0.577\dots$ is the Euler constant.\\

Eqs. (\ref{jseries}, \ref{yseries}) represent the Riccati-Bessel and
Riccati-Neuman functions in the form of infinite series. Each term of these
series is a product of a function depending on $k$ and another function
depending on $r$, i.e. the $k$ and $r$ dependencies are given in a separable
form.\\

What do the above formulae tell us about the Jost functions? The functions
$j_\lambda(kr)$ and $y_\lambda(kr)$ are involved in the coefficients of the
differential equations (\ref{eqA},\ref{eqB}) that determine the Jost functions.
This means that the
Jost functions are not single valued functions of the energy.
Indeed, for each
choice of $E$, we have two possible values of the momentum
$$
     k=\pm\sqrt{\frac{2\mu E}{\hbar^2}}\ .
$$
The index $\lambda=-1/2,\,1/2,\,3/2,\,\dots$ of the Riccati functions is
a half-integer. This means that the differential equations
involve such multi-valued functions as square-root and logarithm of the
momentum.\\

Therefore the Jost functions are
defined on a complicated Riemann surface and the threshold point $E=0$ is a
branching point of this surface. It would be desirable to find an expression
for the Jost functions in terms of the powers of $\sqrt{k}$, the logarithmic
function $h(k)$, and some entire single valued functions of $E$. In order to do
this, we notice that the series in Eqs. (\ref{jseries},\ref{yseries}) involve
only even powers of $k$, i.e. the powers of the energy $k^2=2\mu E/\hbar^2$.
Since for any finite $r$ these series are absolutely and uniformly convergent on
the whole complex plane of $E$, they define some entire functions, i.e.
\begin{eqnarray}
\label{jfact}
    j_\lambda(kr) &=&
    k^{\lambda+1}\tilde{j}_\lambda(E,r)\ ,\\[3mm]
\label{yfact}
    y_\lambda(kr) &=&
    k^{-\lambda}\tilde{y}_\lambda(E,r)+
    k^{\lambda+1}h(k)\tilde{j}_\lambda(E,r)\ ,
\end{eqnarray}
where the "tilded" functions
\begin{eqnarray}
\label{jtilde}
    \tilde{j}_\lambda(E,r) &=&
    \sum_{n=0}^\infty \left(\frac{2\mu E}{\hbar^2}\right)^n
    f_n^{(\lambda)}(r)\ ,\\[3mm]
\label{ytilde}
    \tilde{y}_\lambda(E,r) &=&
    \sum_{n=0}^\infty \left(\frac{2\mu E}{\hbar^2}\right)^n
    g_n^{(\lambda)}(r)\ ,
\end{eqnarray}
are single-valued entire functions of complex variable $E$.\\

Let us find a similar structure for the functions $A_\ell(E,r)$ and
$B_\ell(E,r)$ and through them for $F_\ell^{\mathrm{(in/out)}}(E,r)$. For this,
we replace the set of equations (\ref{eqA},\ref{eqB}) with their linear
combinations. Namely, we multiply Eq.~(\ref{eqB}) by $h(k)$ and subtract the
result from Eq.~(\ref{eqA}); and as the second equation, we take
Eq.~(\ref{eqB}) multiplied by $k^{-(2\lambda+1)}$. As a result, we obtain:
\begin{eqnarray}
\label{eqA1}
   \partial_r(A_\ell-hB_\ell) &=&
   -\frac{1}{k}(y_\lambda-hj_\lambda) V(A_\ell j_\lambda-
   B_\ell y_\lambda)\ ,\\[3mm]
\label{eqB1}
   \partial_rk^{-(2\lambda+1)}B_\ell &=&
   -k^{-2(\lambda+1)}j_\lambda V(A_\ell j_\lambda-
   B_\ell y_\lambda)\ .
\end{eqnarray}
Now, taking into account Eqs. (\ref{jfact},~\ref{yfact}), we see that
$$
   y_\lambda-hj_\lambda=k^{-\lambda}\tilde{y}_\lambda+
   k^{\lambda+1}h\tilde{j}_\lambda-k^{\lambda+1}h\tilde{j}_\lambda=
   k^{-\lambda}\tilde{y}_\lambda
$$
and
\begin{eqnarray*}
   A_\ell j_\lambda-B_\ell y_\lambda &=&
   A_\ell k^{\lambda+1}\tilde{j}_\lambda-B_\ell k^{-\lambda}\tilde{y}_\lambda-
   B_\ell k^{\lambda+1}h\tilde{j}_\lambda\\[3mm]
   &=&
   k^{\lambda+1}(A_\ell-hB_\ell)\tilde{j}_\lambda-k^{-\lambda}
   B_\ell\tilde{y}_\lambda\ .
\end{eqnarray*}
Substituting these expressions into Eqs. (\ref{eqA1},~\ref{eqB1}), we have
\begin{eqnarray}
\label{eqA2}
   \partial_r(A_\ell-hB_\ell) &=&
   -k^{-(\lambda+1)}\tilde{y}_\lambda V\left[
    k^{\lambda+1}(A_\ell-hB_\ell)\tilde{j}_\lambda
    -k^{-\lambda}B_\ell \tilde{y}_\lambda
    \right]\ ,\\[3mm]
\label{eqB2}
   \partial_rk^{-(2\lambda+1)}B_\ell &=&
   -k^{-(\lambda+1)}\tilde{j}_\lambda V\left[
    k^{\lambda+1}(A_\ell-hB_\ell)\tilde{j}_\lambda
    -k^{-\lambda}B_\ell \tilde{y}_\lambda
    \right]\ .
\end{eqnarray}
If we introduce the "tilded" functions
\begin{eqnarray}
\label{Atilde}
   \tilde{A}_\ell(E,r) &\equiv& A_\ell(E,r)-h(k)B_\ell(E,k)\ ,\\[3mm]
   \tilde{B}_\ell(E,r) &\equiv& k^{-(2\lambda+1)}B_\ell(E,r)\ ,
\end{eqnarray}
then Eqs. (\ref{eqA2},~\ref{eqB2}) assume the following form
\begin{eqnarray}
\label{eqAtilde}
   \partial_r\tilde{A}_\ell &=&
   -\tilde{y}_\lambda V\left(
    \tilde{A}_\ell\tilde{j}_\lambda-\tilde{B}_\ell\tilde{y}_\lambda
    \right)\ ,\\[3mm]
\label{eqBtilde}
   \partial_r\tilde{B}_\ell &=&
   -\tilde{j}_\lambda V\left(
    \tilde{A}_\ell\tilde{j}_\lambda-\tilde{B}_\ell\tilde{y}_\lambda
    \right)
\end{eqnarray}
with the boundary conditions [that follow from (\ref{bcondAB})]
\begin{equation}
\label{bcondtilde}
    \tilde{A}_\ell(E,0)=1\ ,\qquad
    \tilde{B}_\ell(E,0)=0\ .
\end{equation}
For any finite $r$, all the coefficient functions in Eqs.
(\ref{eqAtilde},~\ref{eqBtilde}) are
entire functions of the parameter $E$ and the boundary conditions are
$E$-independent. According to the Poincar\'e theorem \cite{Poincare} the
solutions of these
equations, i.e. the functions $\tilde{A}_\ell(E,r)$ and $\tilde{B}_\ell(E,r)$,
are entire (analytic single-valued) functions of the complex variable $E$.\\

Therefore the structure we wanted to find is as follows:
\begin{eqnarray}
\label{Afactorized}
   A_\ell(E,r) &=& \tilde{A}_\ell(E,r)+k^{2\lambda+1}h(k)\tilde{B}_\ell(E,r)
   \ ,\\[3mm]
\label{Bfactorized}
   B_\ell(E,r) &=& k^{2\lambda+1}\tilde{B}_\ell(E,r)\ ,
\end{eqnarray}
where $\tilde{A}_\ell(E,r)$ and $\tilde{B}_\ell(E,r)$ are single-valued
analytic functions of $E$. Apart from these single-valued functions, the
original functions $A_\ell$ and $B_\ell$ involve the factors $k^{2\lambda+1}$
and $h(k)$. Since $\lambda$ is half-integer, the power $(2\lambda+1)$ is always
even and thus $k^{2\lambda+1}$ is also single-valued function of $E$, but $h(k)$
has a logarithmic branching point at $E=0$.\\

The functions $F_\ell^{\mathrm{(in/out)}}$ have similar structure
\begin{eqnarray}
\label{Finfactorized}
   F_\ell^{\mathrm{(in)}}(E,r)
   &=&
   \frac12(A_\ell-iB_\ell)=\frac12\left\{
   \tilde{A}_\ell(E,r)+k^{2\lambda+1}[h(k)-i]\tilde{B}_\ell(E,r)\right\}
   \ ,\\[3mm]
\label{Foutfactorized}
   F_\ell^{\mathrm{(out)}}(E,r)
   &=&
   \frac12(A_\ell+iB_\ell)=\frac12\left\{
   \tilde{A}_\ell(E,r)+k^{2\lambda+1}[h(k)+i]\tilde{B}_\ell(E,r)\right\}
   \ .
\end{eqnarray}
\section{Analytic structure of the Jost functions}
\label{sec.Analytic_structure}
What we have established in the previous Section, is valid for any complex
$E$ and any finite distance $r$. In other words, so far we have established
that the Jost functions (\ref{liminout}) have the structure
(\ref{Finfactorized}, \ref{Foutfactorized}) if the potential is cut off at
certain radius $r=R$ (does not matter how large $R$ is). The problem is that to
prove analyticity of $\tilde{A}_\ell(E,r)$ and $\tilde{B}_\ell(E,r)$ with
respect to variable $E$, we used the Poincar\'e theorem which requires that the
coefficients of Eqs. (\ref{eqAtilde},\ref{eqBtilde}) be holomorphic functions of
$E$. If $E$ is a real positive number and the potential is of a short range,
this is true even if $r\to\infty$. Indeed, in such a case both
$\tilde{j}_\lambda(kr)$ and $\tilde{y}_\lambda(kr)$ oscillate with
finite amplitudes even at infinity and thus the coefficients of Eqs.
(\ref{eqAtilde},\ref{eqBtilde}) simply tend to zero (i.e. remain holomorphic)
when $r\to\infty$. If however $E$ is negative or complex, then generally
speaking this is not true. As we will see shortly, there still is a domain of
complex $E$ where the coefficients remain holomorphic. In other words, if we
extend $r$ to infinity, we have to narrow the domain of $E$.\\

The Riccati-Bessel and Riccati-Neumann functions are linear
combinations of the Riccat-Hankel functions and thus at large distances behave
as exponential functions (\ref{hH}). If the momentum
has a nonzero imaginary part, then one or the other of these exponentials is
diverging and thus both $\tilde{j}_\lambda(kr)$ and $\tilde{y}_\lambda(kr)$
tend to infinity when $r\to\infty$.
To some extent the situation can be saved by using a short-range (exponentially
decaying) potential  $V(r)\sim\exp(-\eta r)$,
which compensates the divergence of $\tilde{j}_\lambda(kr)$ and
$\tilde{y}_\lambda(kr)$ within
certain domain ${\cal D}$ of the complex $E$-plane along its real axis. The
borders of the domain ${\cal D}$
are determined by the requirement that none of the coefficients of
Eqs. (\ref{eqAtilde},\ref{eqBtilde})
are divergent. The behavior (convergent or divergent) of these coefficients is
determined by the
product $\exp(\pm 2ikr)\exp(-\eta r)$. For a given $\eta$ it is not difficult to
find the domain ${\cal D}$,
\begin{equation}
\label{domainD}
   {\cal D}=\left\{
   E: \left|2\mathrm{Im}\,\sqrt{2\mu E/\hbar^2}\right|<\eta\right\}\ ,
\end{equation}
which gives the condition
\begin{equation}
\label{domainD_condition}
   \left(\mathrm{Im}\,E\right)^2<\frac{\hbar^4\eta^4}{16\mu^2}+
   \frac{\hbar^2\eta^2}{2\mu}\mathrm{Re}\,E\ .
\end{equation}
Similar analyticity domain was obtained by
Motovilov \cite{Motovilov} for the three-dimensional multi-channel $T$-matrix,
using a rigorous analysis of the corresponding scattering operators.\\

\begin{figure}[ht!]
\centerline{\epsfig{file=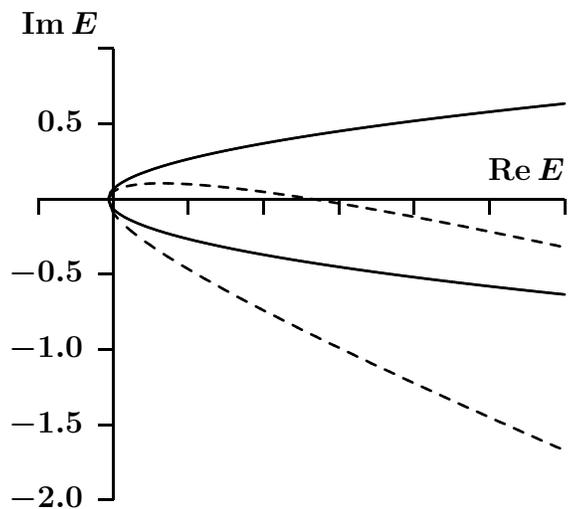}}
\caption{\sf
Domains $\cal D$ for the potential (\ref{Example.V}), defined by Eq.
(\ref{domainD_condition}) (within the solid
curve) and by Eq. (\ref{domainDkr_condition}) for the rotation angle
$\theta=0.05\pi$ (within the dashed curve). The energy is given in the
donor Hartree units (see Sec. \ref{sec.Example}).}
\label{fig.domainD}
\end{figure}

The faster the potential decays, the wider is the domain. An example of such a
domain is shown in figure \ref{fig.domainD} for the model used in section
\ref{sec.Example}. It is a parabolic domain along the real axis, whose border is
shown by the solid curve. It crosses the real axis at $E\approx-0.0332$ (in
donor Hartree units). When $r\to\infty$, we can only use the Poincar\'e theorem
within ${\cal D}$, and thus we can say that at least within this domain the
functions $\tilde{A}_\ell(E,\infty)$ and $\tilde{B}_\ell(E,\infty)$ are the
holomorphic functions of variable $E$.\\

If the potential $V(r)$ (or at least its long-range tail) is an analytic
function of complex variable $r$ and exponentially decays along any ray
$r=|r|e^{i\theta}$ within certain sector of the complex $r$-plane, then the
domain ${\cal D}$ can be extended by using the complex rotation described in
section \ref{sec.complex_rotation}. With a complex radius, the product
$\exp(\pm 2ikr)\exp(-\eta r)$ vanishes at infinity if $E$ is within
\begin{equation}
\label{domainDkr}
   {\cal D}=\left\{
   E: \left|2\mathrm{Im}\,(kr)\right|<\eta\mathrm{Re}\,r\right\}\ ,
\end{equation}
which generalizes Eq. (\ref{domainD}). It is easy to show that if
$E=|E|\exp(i\chi)$ then such a domain can be defined by the following inequality
\begin{equation}
\label{domainDkr_condition}
   \sin^2\left(\frac{\chi}{2}+\theta\right)<
   \frac{\hbar^2\eta^2\cos^2\theta}{8\mu|E|}\ .
\end{equation}
With $\theta=0$ this condition is transformed
into (\ref{domainD_condition}). An example of such a domain with the rotation
angle $\theta=0.05\pi$ for the potential (\ref{Example.V}) is shown by the
dashed curve in Fig. \ref{fig.domainD}.\\

The physically interesting domain of the $E$-plane where the structure
(\ref{Finfactorized}, \ref{Foutfactorized}), can be used in
practical calculations, lies on the positive real axis (scattering) and in the
close vicinity below it (pronounced resonances). Therefore, we can say that to
all practical purposes this structure is valid at an arbitrary point $E$.\\

If we denote the asymptotic values (within ${\cal D}$) of the "tilded" functions
as
\begin{equation}
\label{aA}
   \tilde{a}_\ell(E)=\lim_{r\to\infty}\tilde{A}_\ell(E,r)\ ,
   \qquad
   \tilde{b}_\ell(E)=\lim_{r\to\infty}\tilde{B}_\ell(E,r)\ ,
\end{equation}
Then the Jost functions and the $S$-matrix can be written as follows
\begin{eqnarray}
\label{Jinfactorized}
   f_\ell^{\mathrm{(in)}}(E)
   &=&
   \frac12\left\{
   \tilde{a}_\ell(E)+k^{2\lambda+1}[h(k)-i]\tilde{b}_\ell(E)\right\}
   \ ,\\[3mm]
\label{Joutfactorized}
   f_\ell^{\mathrm{(out)}}(E)
   &=&
   \frac12\left\{
   \tilde{a}_\ell(E)+k^{2\lambda+1}[h(k)+i]\tilde{b}_\ell(E)\right\}
   \ ,
\end{eqnarray}
\begin{equation}
\label{sfactorized}
   s_\ell(E)=
   \frac{\tilde{a}_\ell(E)+k^{2\lambda+1}[h(k)+i]\tilde{b}_\ell(E)}
        {\tilde{a}_\ell(E)+k^{2\lambda+1}[h(k)-i]\tilde{b}_\ell(E)}\ .
\end{equation}
To find the Jost functions or the $S$-matrix on any sheet of the Riemann
surface, we need to calculate the functions $\tilde{a}_\ell(E)$ and
$\tilde{b}_\ell(E)$ only once (because they are single valued). The choice of
the sheet is determined by an appropriate choice of the value of the logarithmic
function $h(k)$. Please note that $k^{2\lambda+1}=k^{2\ell}$ is a
single-valued function of $E$.

\section{Power-series expansions of the Jost functions}
\label{sec.JFexpansions}
The functions $\tilde{a}_\ell(E)$ and $\tilde{b}_\ell(E)$ are holomorphic (i.e.
single-valued and analytic) and therefore can be expanded in Taylor series near
any point $E_0$ within the domain $\cal D$ of the complex energy plane. The
expansion around the point $E_0=0$ will give us the standard effective-range
series. But we can also do such an expansion near an arbitrary point,
\begin{eqnarray}
\label{aexpansion}
    \tilde{a}_\ell(E) &=&
    \sum_{n=0}^\infty \alpha_n^{(\ell)}(E_0)(E-E_0)^n\ ,\\[3mm]
\label{bexpansion}
    \tilde{b}_\ell(E) &=&
    \sum_{n=0}^\infty \beta_n^{(\ell)}(E_0)(E-E_0)^n\ .
\end{eqnarray}
How the expansion coefficients $\alpha_n^{(\ell)}$ and $\beta_n^{(\ell)}$ can
be found? For this purpose, we can derive differential equations, the solutions
of which asymptotically tend to $\alpha_n^{(\ell)}$ and $\beta_n^{(\ell)}$.
Indeed, such an expansion can be done at any fixed radius $r$ because the
functions $\tilde{A}_\ell(E,r)$ and $\tilde{B}_\ell(E,r)$ reach their limits
(\ref{aA}) at $r$ if the potential is cut off at this radius (in the spirit of
the variable-phase approach). Therefore for each $r$, we have
\begin{eqnarray}
\label{Aexp}
    \tilde{A}_\ell(E,r) &=&
    \sum_{n=0}^\infty {\cal A}_n^{(\ell)}(E_0,r)(E-E_0)^n\ ,\\[3mm]
\label{Bexp}
    \tilde{B}_\ell(E,r) &=&
    \sum_{n=0}^\infty {\cal B}_n^{(\ell)}(E_0,r)(E-E_0)^n\ ,
\end{eqnarray}
where
\begin{equation}
\label{Aalpha}
   \alpha_n^{(\ell)}(E_0)=\lim_{r\to\infty}{\cal A}_n^{(\ell)}(E_0,r)\ ,
   \qquad
   \beta_n^{(\ell)}(E_0)=\lim_{r\to\infty}{\cal B}_n^{(\ell)}(E_0,r)\ .
\end{equation}
Therefore, the differential equations mentioned above, should determine the
functions ${\cal A}_n^{(\ell)}(E_0,r)$ and ${\cal B}_n^{(\ell)}(E_0,r)$. In order to
obtain such equations, we expand the "tilded" functions
$\tilde{j}_\lambda(E,r)$ and $\tilde{y}_\lambda(E,r)$ in the Taylor series near
an arbitary point $E_0$
\begin{eqnarray}
\label{jtildeE0}
    \tilde{j}_\lambda(E,r) &=&
    \sum_{n=0}^\infty
    s_n^{(\lambda)}(E_0,r)\left(E-E_0\right)^n\ ,\\[3mm]
\label{ytildeE0}
    \tilde{y}_\lambda(E,r) &=&
    \sum_{n=0}^\infty
    c_n^{(\lambda)}(E_0,r)\left(E-E_0\right)^n\ ,
\end{eqnarray}
which are more general expansions than the series (\ref{jtilde},\ref{ytilde})
for the particular case of the threshold energy $E_0=0$. Any number of the
expansion coefficients $s_n^{(\lambda)}(E_0,r)$ and $c_n^{(\lambda)}(E_0,r)$
can be found using the recurrence relations derived in the Appendix
\ref{app.coefficients}.\\

Substituting the expansions
(\ref{Aexp},\ref{Bexp},\ref{jtildeE0},\ref{ytildeE0}) into
Eqs.~(\ref{eqAtilde},\ref{eqBtilde}), and equalizing the factors of the same
powers of $(E-E_0)$, we obtain the equations we are looking for,
\begin{eqnarray}
\label{eqAtildexp}
   \partial_r\tilde{\cal A}_n^{(\ell)} &=&
   -\sum_{i+j+k=n}c_i^{(\lambda)}V
   \left(
    \tilde{\cal A}_j^{(\ell)}s_k^{(\lambda)}-
    \tilde{\cal B}_j^{(\ell)}c_k^{(\lambda)}
    \right)\ ,\\[3mm]
\label{eqBtildexp}
   \partial_r\tilde{\cal B}_n^{(\ell)} &=&
   -\sum_{i+j+k=n}s_i^{(\lambda)}V
   \left(
    \tilde{\cal A}_j^{(\ell)}s_k^{(\lambda)}-
    \tilde{\cal B}_j^{(\ell)}c_k^{(\lambda)}
    \right)\ ,
\end{eqnarray}
with the boundary conditions
\begin{equation}
\label{bcondABexpansion}
   \tilde{\cal A}_n^{(\ell)}(E_0,0)=\delta_{n0}\ ,
   \qquad
   \tilde{\cal B}_n^{(\ell)}(E_0,0)=0\ ,
   \qquad
   n=0,1,2,3,\dots
\end{equation}
These conditions follow from the fact that the corresponding boundary conditions
(\ref{bcondtilde}) do not depend on $E$. Therefore, starting with the initial
values (\ref{bcondABexpansion}) at $r=0$, and numerically solving first $N+1$
pairs of differential equations of the system (\ref{eqAtildexp},
\ref{eqBtildexp}) up to a sufficiently large radius $r_{\mathrm{max}}$, we
obtain first $N+1$ expansion coefficients
\begin{equation}
\label{Ncoefficients}
   \alpha_n^{(\ell)}(E_0)=
   \tilde{\cal A}_n^{(\ell)}(E_0,r_{\mathrm{max}})\ ,
   \qquad
   \beta_n^{(\ell)}(E_0)=
   \tilde{\cal B}_n^{(\ell)}(E_0,r_{\mathrm{max}})\ ,
   \qquad
   n=0,1,2,\dots,N
\end{equation}
These coefficients give us the following approximate formulae for the Jost functions
\begin{eqnarray}
\label{Jinapprox}
   f_\ell^{\mathrm{(in)}}(E)
   &\approx&
   \frac12\sum_{n=0}^N\left\{
   \alpha_n^{(\ell)}(E_0)+k^{2\lambda+1}[h(k)-i]
   \beta_n^{(\ell)}(E_0)\right\}(E-E_0)^n
   \ ,\\[3mm]
\label{Joutapprox}
   f_\ell^{\mathrm{(out)}}(E)
   &\approx&
   \frac12\sum_{n=0}^N\left\{
   \alpha_n^{(\ell)}(E_0)+k^{2\lambda+1}[h(k)+i]
   \beta_n^{(\ell)}(E_0)\right\}(E-E_0)^n
   \ ,
\end{eqnarray}
which are valid for any complex value of $E$ within a domain around the chosen central
point $E_0$. Apparently, the closer $E$ is to $E_0$, the better is the accuracy of these
formulae. It is interesting to note that Eq. (9) of Ref. \cite{Klawunn} is the
first term of our Eq. (\ref{Jinapprox}) for the particular case of $E_0=0$.\\

An alternative way of using formulae (\ref{Jinapprox},\ref{Joutapprox}) is to
treat the expansion coefficients $\alpha_n^{(\ell)}(E_0)$,
$\beta_n^{(\ell)}(E_0)$, $n=0,1,\dots,N$ as fitting parameters. Adjusting them
in such a way that the corresponding cross section (see Appendix
\ref{app.scatt.cross.section}) reproduces experimental data in vicinity of a
real energy $E_0$, one then can use the Jost function
(\ref{Jinapprox}) at the nearby complex energies for locating possible
resonances. The obvious advantage of such an approach is that the resonance
energy and the width are deduced directly from experimental data using correct
analytic structure of the $S$-matrix.

\section{Effective-range expansion}
\label{sec.EffRangeExp}
Far away from the interaction region the radial wave function (\ref{ansatzAB}) is a
linear combination of the Riccati-Bessel and Riccati-Neumann functions
\begin{equation}
\label{uABass}
   u_\ell(E,r)
   \ \mathop{\longrightarrow}\limits_{r\to\infty}
   \ a_\ell(E)j_\lambda(kr)-b_\ell(E)y_\lambda(kr)\ ,
\end{equation}
where
\begin{equation}
\label{abAB}
   a_\ell(E)=\lim_{r\to\infty}A_\ell(E,r)\ ,
   \qquad
   b_\ell(E)=\lim_{r\to\infty}B_\ell(E,r)\ .
\end{equation}
The functions $j_\lambda$ and $y_\lambda$ in (\ref{uABass}) can be written in their
asymptotic form,
\begin{eqnarray}
\label{jass}
   j_\lambda(kr)
   \ &\mathop{\longrightarrow}\limits_{r\to\infty}&
   \ \phantom{+}\sin\left(kr-\frac{\lambda\pi}{2}\right)\ ,\\[3mm]
\label{yass}
   y_\lambda(kr)
   \ &\mathop{\longrightarrow}\limits_{r\to\infty}&
   \ -\cos\left(kr-\frac{\lambda\pi}{2}\right)\ ,
\end{eqnarray}
which gives
\begin{eqnarray*}
   u_\ell(E,r)
   \ \mathop{\longrightarrow}\limits_{r\to\infty}
   \ &&
   a_\ell(E)\sin\left(kr-\frac{\lambda\pi}{2}\right)+
   b_\ell(E)\cos\left(kr-\frac{\lambda\pi}{2}\right)=\\[3mm]
   &&
   =N\sin\left[kr-\frac{\lambda\pi}{2}+\delta_\ell(E)\right]\ ,
\end{eqnarray*}
where $a_\ell$ and $b_\ell$ are replaced with their common normalization factor $N$
and the scattering phase shift $\delta_\ell$,
\begin{eqnarray}
\label{adelta}
   a_\ell(E) &=& N\cos\delta_\ell(E)\ ,\\[3mm]
\label{bdelta}
   b_\ell(E) &=& N\sin\delta_\ell(E)\ .
\end{eqnarray}
Using the relations (\ref{Afactorized}, \ref{Bfactorized}) at large distances
($r\to\infty$),
\begin{eqnarray}
\label{afactor}
   a_\ell(E) &=& \tilde{a}_\ell(E)+k^{2\lambda+1}h(k)\tilde{b}_\ell(E)
   \ ,\\[3mm]
\label{bfactor}
   b_\ell(E) &=& k^{2\lambda+1}\tilde{b}_\ell(E)\ ,
\end{eqnarray}
we can construct the so called effective-range function which is a holomorphic function
of the energy. This is done by taking the ratio
\begin{eqnarray*}
   \cot\delta_\ell &=& \frac{a_\ell}{b_\ell}=
   \frac{\tilde{a}_\ell+k^{2\lambda+1}h\tilde{b}_\ell}
   {k^{2\lambda+1}\tilde{b}_\ell}\ ,
\end{eqnarray*}
and moving all the "troublesome" terms and factors which may generate singularities,
to the left hand side of the equation,
$$
   k^{2\lambda+1}\cot\delta_\ell = \frac{\tilde{a}_\ell}{\tilde{b}_\ell}
   +k^{2\lambda+1}h\ ,
$$
\begin{equation}
\label{EffRangeFunct}
   k^{2\lambda+1}\left[\cot\delta_\ell(E)-h(k)\right]=
   k^{2\ell}\left[\cot\delta_\ell(E)-h(k)\right]
   =
   \frac{\tilde{a}_\ell(E)}{\tilde{b}_\ell(E)}\ .
\end{equation}
Both the numerator and denominator in the last ratio can be written in the form of
power series (\ref{aexpansion}, \ref{bexpansion}) with $E_0=0$
\begin{equation}
\label{ratiopolyn}
   k^{2\lambda+1}\left[\cot\delta_\ell(E)-h(k)\right]
   =
   \frac{\displaystyle\sum_{n=0}^\infty \alpha_n^{(\ell)}(0)E^n}
   {\displaystyle\sum_{n=0}^\infty \beta_n^{(\ell)}(0)E^n}\ .
\end{equation}
Using Eq. (3.6.22) of the book by Abramowitz et al.,
$$
   \frac{a_0+a_1x+a_2x^2+\cdots}{b_0+b_1x+b_2x^2+\cdots}=
   \frac{a_0}{b_0}\left[
   1+\left(\frac{a_1}{a_0}-\frac{b_1}{b_0}\right)x+
   \left(\frac{a_2}{a_0}-
   \frac{b_1(a_1b_0-a_0b_1)}{a_0b_0^2}-\frac{b_2}{b_0}
   \right)x^2+\cdots\right]\ ,
$$
the division of two polynomials in Eq. (\ref{ratiopolyn}) is done as follows
\begin{equation}
\label{Eff_range_formula}
   k^{2\lambda+1}\left[\cot\delta_\ell(E)-h(k)\right]
   =
   -\frac{1}{a^{(\ell)}}+\frac{r_0^{(\ell)}}{2}k^2+\cdots\ ,
\end{equation}
where the scattering length $a^{(\ell)}$ and the effective radius $r_0^{(\ell)}$ for
the state with the angular momentum $\ell$ are given by
\begin{eqnarray}
\label{sclength}
   a^{(\ell)} &=& -\frac{\beta_0^{(\ell)}}{\alpha_0^{(\ell)}}\ ,\\[3mm]
\label{effectiveradius}
   r_0^{(\ell)} &=& \frac{\hbar^2}{\mu}\left(
   \frac{\alpha_1^{(\ell)}}{\beta_0^{(\ell)}}-
   \frac{\alpha_0^{(\ell)}\beta_1^{(\ell)}}{\beta_0^{(\ell)2}}\right)\ .
\end{eqnarray}

\section{A numerical example related to quantum dot theory}
\label{sec.Example}
To demonstarate how the proposed method works, we use the following
circularly-symmetric potential, which is motivated by the models that are
currently used in the theory of quantum dots,
\begin{equation}
\label{Example.V}
    U(r)=V_0(r-r_0)e^{-r/R}\ ,
\end{equation}
with $V_0=25$, $r_0=2$, and $R=2$, where $V_0$ (as well as all the energies in
this example) is measured in the so called "donor Hartree units" and the
distances in the units of "donor Bohr radius", which were chosen to be
$10.96\,\mathrm{meV}$ and $101.89\,\mathrm{\AA}$, respectively. These values for
the units are relevant to the motion of electrons in the semiconductor material
GaAs \cite{Markvoort}, where the effective electron mass is $\mu=0.063m_e$ (with
$m_e$ being free electron mass).\\

\begin{figure}[ht!]
\centerline{\epsfig{file=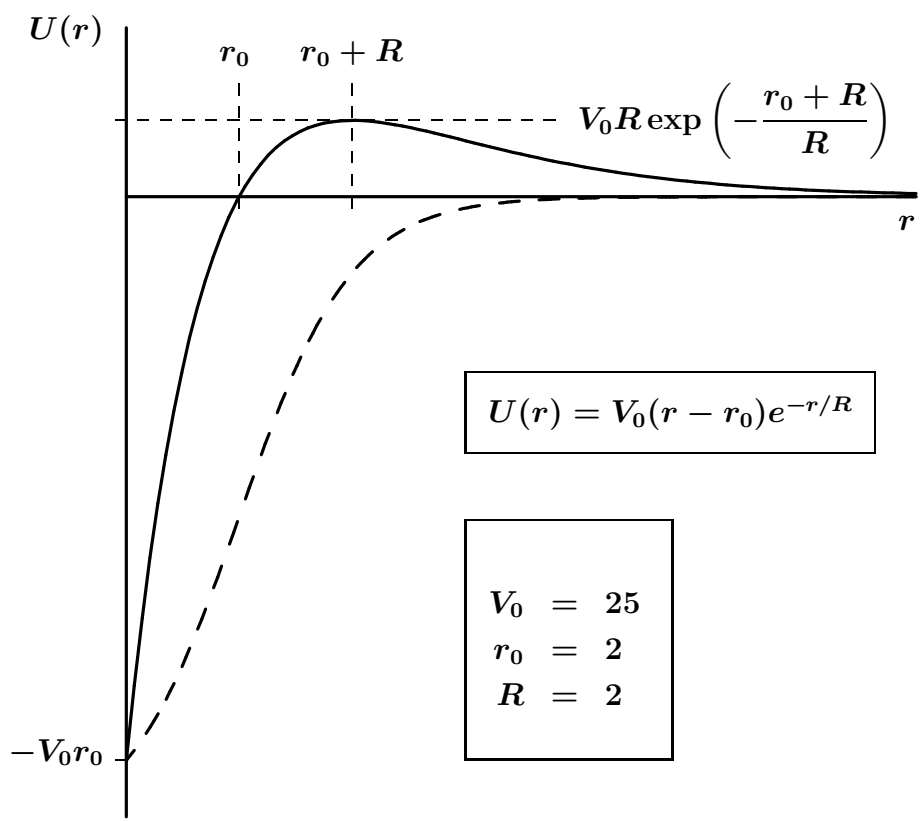}}
\caption{\sf
Model potential (\ref{Example.V}) measured in the
donor Hartree units $10.96\,\mathrm{meV}$ as a function of the distance measured
in the units of donor Bohr radius $101.89\,\mathrm{\AA}$. The dashed curve is a
typical potential for an empty two-dimensional quantum dot.
}
\label{fig.potential}
\end{figure}

Although, strictly, the potential (\ref{Example.V}) should be considered as
an abstract
quantum-mechanical "toy" model, we chose its shape in such a way that it
resembles the potentials that are currently used to describe
two-dimensional
quantum dots (see, for example, Refs. \cite{Markvoort, Filippo, Adamowski,
Ciftja}). As is seen in Fig.\ref{fig.potential}, our potential has a repulsive
barrier which is not present in the traditional quantum-dot models. The main
reason for introducing such a barrier was to enrich our "toy" model spectrum
with resonances. However, one can argue that such a barrier may appear in real
quantum dots as well. Indeed, when electrons fill up the lower levels of a dot,
they should repel each other and tend to stay mostly at its periphery. This
means that for an additional incoming electron the attractive force at the
centre is reduced and a repulsion appears at the border. In other words, the
original empty-dot confining-potential (shown with the dashed curve) is
transformed into something that looks like our "toy" potential. Of course, this
speculative reasoning does not mean that we claim that our potential is anything
more than an abstract model.\\

\begin{figure}[ht!]
\centerline{\epsfig{file=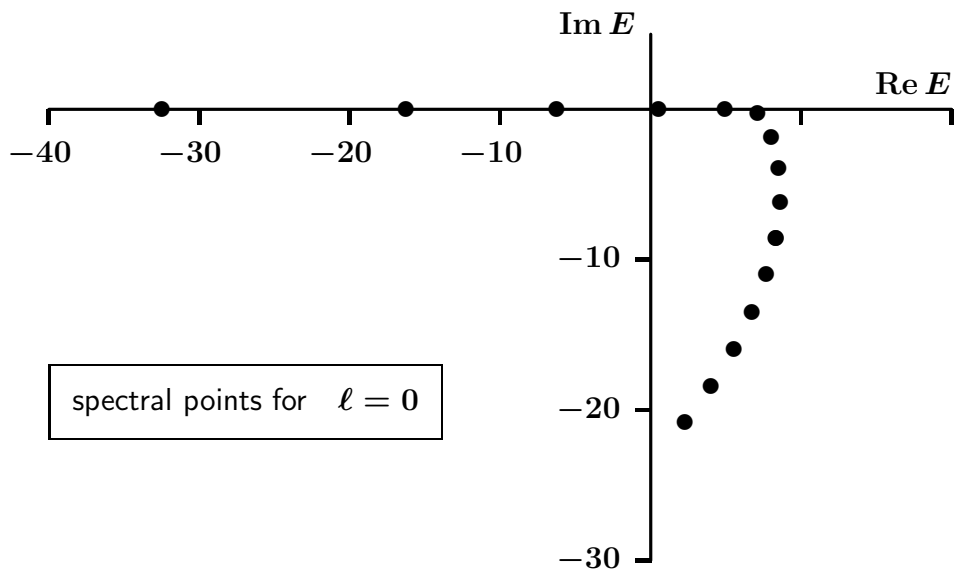}}
\caption{\sf
Spectral points generated by the potential (\ref{Example.V}). Their numerical
values are given in Table~\protect\ref{table.spectrum}.
}
\label{fig.spectrum}
\end{figure}

Since nothing special is associated with the angular momentum, we only consider
here the $S$-wave states ($\ell=0$). For such a case, the potential
(\ref{Example.V}) supports three bound states and a squence of resonances. These
spectral points (given in Table \ref{table.spectrum} and shown in
Fig.~\ref{fig.spectrum}) were located using the exact approach, i.e. as the
roots of Eq. (\ref{spectralEQ}), where $f_\ell^{(\mathrm{in})}$ is the
asymptotic value (\ref{liminout}) of the solution of Eq. (\ref{eqFin}).\\

\begin{figure}[ht!]
\centerline{\epsfig{file=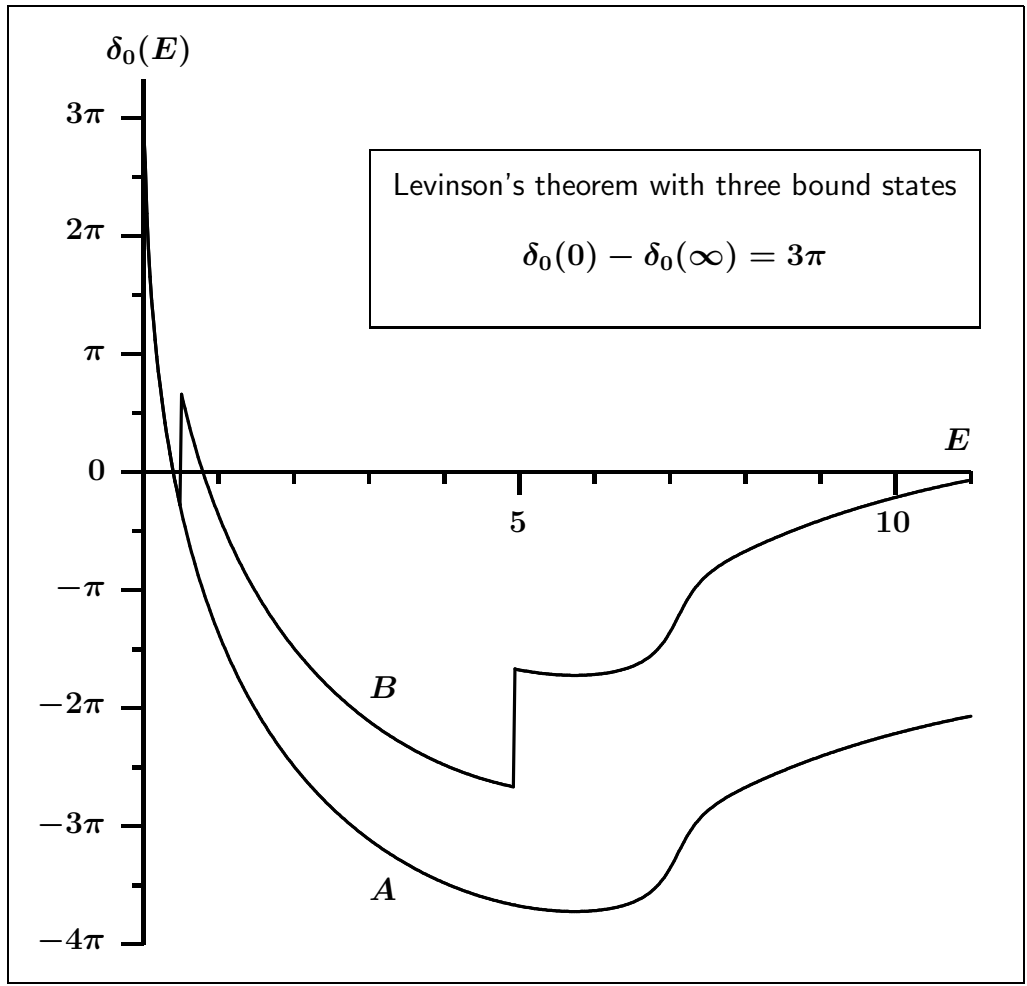}}
\caption{\sf
$S$-wave scattering phase-shift for the potential (\ref{Example.V}).
In the curve "A" the sharp jumps in $\pi$ (corresponding to the first two
extremely narrow resonances) are missing and as a result it does not obey the
Levinson's theorem.
}
\label{fig.phase_shift}
\end{figure}

To make sure that we did not miss any of the bound states and/or narrow resonances, we
calculated the $S$-wave scattering phase-shift and checked if it obeys the Levinson's
theorem. In Refs.\cite{Gibson, Cheney, Bolle3} it was shown that in the absence
of a zero-energy bound state for the $P$-wave and always for the $S$-wave, this
theorem is the same as for the three-dimensional scattering, namely,
\begin{equation}
\label{Levinson_theorem}
      \delta_\ell(0)-\delta_\ell(\infty)=\pi N_\ell\ ,
\end{equation}
where $N_\ell$ is the number of bound states with the angular momentum $\ell$. If the
energy moves to the right along the real axis, the phase shift increases by $\pi$ near
each resonance which is not far from the real axis. The smaller is the width, the more
sharp is the increase. When calculating the phase shift numerically, it is easy to miss a
sharp jump corresponding a narrow resonance. The curve "A" in Fig. \ref{fig.phase_shift}
is an example of such omissions (the first two resonances are missed because of a too
large step along the $E$-axis). The correct phase shift is shown by the curve "B". It
starts with $3\pi$ at the threshold and tends to zero at the infinity, in accordance with
Eq. (\ref{Levinson_theorem}).\\

\begin{figure}[ht!]
\centerline{\epsfig{file=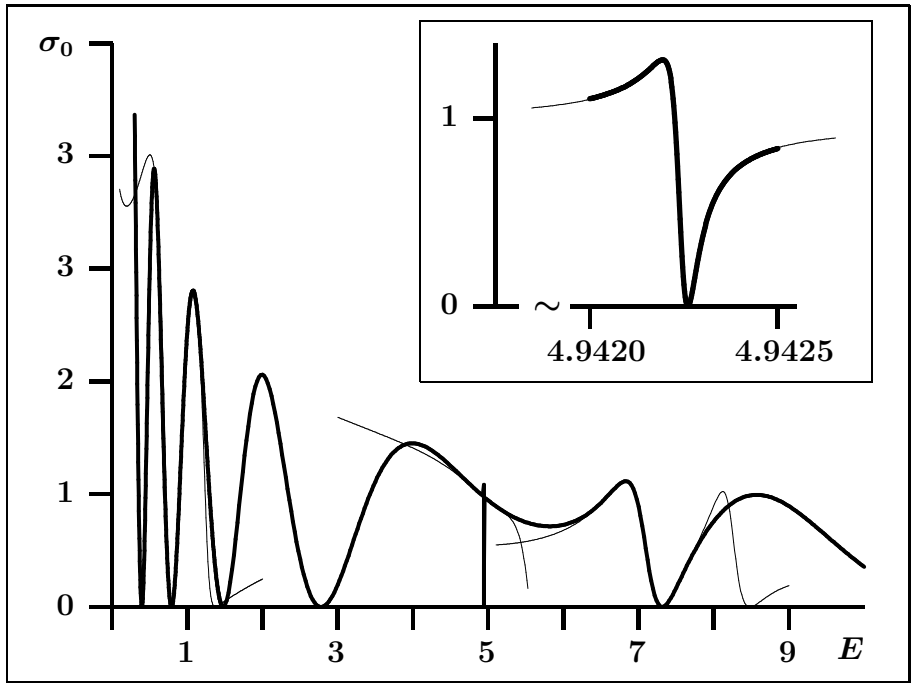}}
\caption{\sf
Comparison of the exact (thick curve) and approximate (thin curve) $S$-wave
cross section for the potential (\ref{Example.V}). The approximate curves are
obtained using the expansion (\ref{Jinapprox}) with $N=4$ (five terms) near the
points $E_0=1\times10.96\,\mathrm{meV}$, $E_0=5\times10.96\,\mathrm{meV}$, and
$E_0=7\times10.96\, \mathrm {meV}$. The insert shows a magnified fragment of
the curves near the second resonance.
}
\label{fig.sigma_5}
\end{figure}

Calculating the first two expansion coefficients and using
Eqs. (\ref{sclength},\ref{effectiveradius}), we found the following scattering length and
effective radius,
$$
    a_0=-0.4521260323\,\mathrm{[dimensionless]}\ ,
    \qquad
    r_0=0.0586790752\,\mathrm{[length^2]}\ .
$$
As a first test of the expansions (\ref{Jinapprox},\ref{Joutapprox}), we
performed them at several scattering energies (i.e. on the real energy axis)
and compared the approximate cross section obtained from the approximate Jost
functions (see Appendix \ref{app.scatt.cross.section}) with the corresponding
exact cross section that was calculated using the exact Jost functions via
numerical integration of the system of differential equations
(\ref{eqFin},\ref{eqFout}). Fig. \ref{fig.sigma_5} shows the exact cross section
in the interval $E\in(0,10]$ (thick curve) and the approximate cross sections
(thin curves) when only the first five terms of the series
(\ref{Jinapprox},\ref{Joutapprox}) were taken into account for $E_0$ being $1$,
$5$, and $7$. It is seen that within rather wide interval around each $E_0$ the
expansion reproduces the cross section very well even with all its zigzags.\\

\begin{figure}[ht!]
\centerline{\epsfig{file=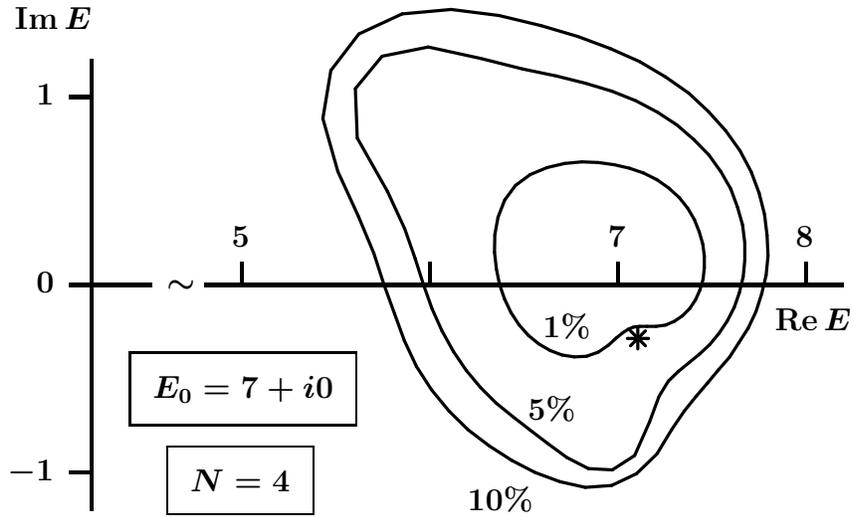}}
\caption{\sf
The domains within which the Jost function
for the potential (\ref{Example.V})
is reproduced, using the first five terms ($N=4$) of the expansion
(\ref{Jinapprox}), with the accuracy better than 1\%,
5\% and 10\%. The expansion is done around the point
$E_0=7\,\times[10.96\,\mathrm{meV}]$ on the real axis. The star shows the third
resonance given in Table~\protect\ref{table.spectrum}.
}
\label{fig.domain5}
\end{figure}

The next step was to test our expansions at complex energies. To begin with,
we performed them around a
point on the real axis, namely, around $E_0=7$ (far away from the threshold
energy) and looked at the Jost function at the nearby complex energies. Why 7?
Simply because there is a resonance not far from this point
(third resonance of Table \ref{table.spectrum}). To check the accuracy of the
expansion, we compared the approximate values of $f_\ell^{\mathrm{(in)}}(E)$ at
various points around $E_0$ with the corresponding exact values of the Jost
function. Apparently, the closer the point $E$ is
to the center of the expansion, the more accurate should be the result. Fig.
\ref{fig.domain5} shows three closed contours around $E_0=7$. Within the
smallest of them the relative error of $f_\ell^{\mathrm{(in)}}(E)$ obtained
by the expansion (\ref{Jinapprox}) with $N=4$ is less than 1\%. The other two
contours show the domains of 5\% and 10\% accuracy. The important fact is that
even if the expansion is done on the real axis, the semi-analytic formulae
(\ref{Jinapprox},\ref{Joutapprox}) remain valid at the nearby complex points.\\

The star in Fig. \ref{fig.domain5} is a resonant zero of the exact Jost
function. As is seen, the 1\%-contour has a "dent" near this point. The reason
for it is that in calculating the relative error, we have an exact value of
$f_\ell^{\mathrm{(in)}}(E)$ in the denominator and this value is zero at the
resonance. By the way, the approximate Jost function (\ref{Jinapprox})
with $N=4$ has zero at
$E=7.1051679246-\frac{i}{2}0.5683685515$
which is not far from its exact position. This means that the
expansion done on the real axis can be used for locating narrow resonances.\\

Finally, we tested the expansion around a point in the fourth quadrant (where
the resonances are) of the complex energy plane. When solving the differential
equations (\ref{eqAtildexp},\ref{eqAtildexp}) we used the complex rotation of
the coordinate (see Sec.\ref{sec.complex_rotation}) with such an angle $\theta$
that $\mathrm{Im}\,(k_0r)=0$ (where $k_0$ is the momentum corresponding to
$E_0$). This guarantees that $E_0$ is within the domain $\cal D$
(see Sec.\ref{sec.Analytic_structure}).\\

\begin{figure}[ht!]
\centerline{\epsfig{file=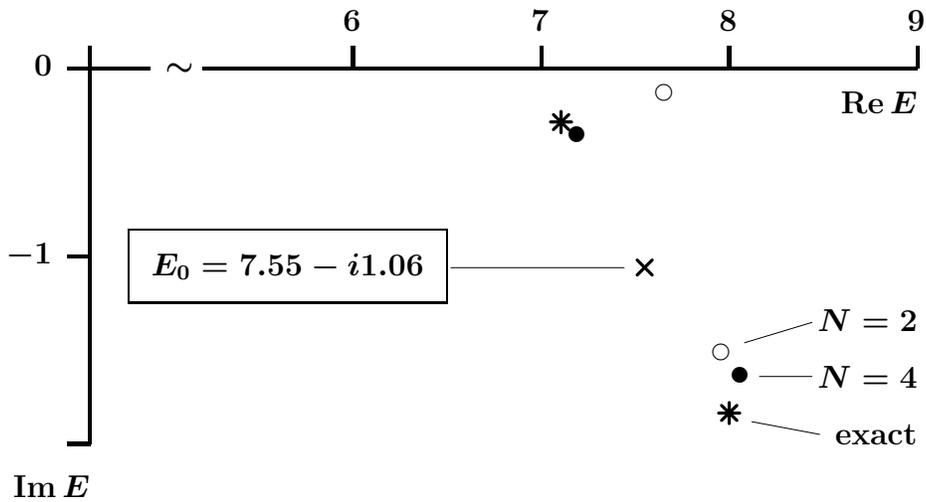}}
\caption{\sf
Spectral points corresponding to the third and fourth resonances of the
potential (\ref{Example.V}). Stars show their exact locations. Open and filled
circles are obtained using the expansion (\ref{Jinapprox}) with $N=2$ (three
terms) and $N=4$ (five terms), respectively. The expansion is done around the
point $E_0=(7.55-i1.06)\times10.96\,\mathrm{meV}$, which is in the middle
between these two resonances.
}
\label{fig.two_resonances}
\end{figure}

Fig. \ref{fig.two_resonances} shows the exact positions of two resonances
(indicated with stars), the center of the expansion (cross) at $E_0=7.55-i1.06$
which is in the middle between them, and two pairs of the approximate locations
of these resonances: open circles for three terms of the expansion and filled
circles for five expansion terms. It is seen that the expansion converges, i.e.
the more terms are taken into account, the more accurately the resonances are
reproduced. It should be noted that the chosen position of $E_0$ is the "worst
case". If we move $E_0$ a bit closer towards one of the resonances, it is
reproduced much more accurately.\\

\section{Conclusion}
\label{sec.Conclusion}
In this paper, we show that the Jost function for the two-dimension scattering
can be written as a sum of two terms, one of which is an analytic single-valued
function of the energy $E$ while the other term can be factorized in  an
analytic function of $E$ and a logarithmic function of the momentum. This means
that the (logarithmic) branching point of the Riemann energy-surface is given in
the Jost function explicitly via the logarithmic factor. The remaining
energy-dependent functions are defined on single energy plane which does not
have any branching points anymore. For these energy-dependent functions, we
derive a system of first-order differential equations. Then, using the fact that
the functions are analytic within certain domain $\cal D$, we expand them in the
power series around an arbitrary point $E_0\in{\cal D}$ and obtain a system of
differential equations that determine the expansion coefficients.\\

A systematic procedure developed in this paper, allows us to accurately
calculate the power series expansion of the Jost function practically at any
point on the Riemann surface of the energy. Actually, the expansion is done for
the single-valued functions of the energy, while the choice of the sheet of the
Riemann surface is done by appropriately choosing the sheet of the logarithmic
function of the momentum.\\

The method suggested in this paper, makes it
possible to obtain a semi-analytic expression for the two-dimensional Jost
function (and therefore for the corresponding $S$-matrix) near an arbitrary
point on the Riemann surface and thus to locate the resonant states as the
$S$-matrix poles. Alternatively, the expansion can be used to parametrize
experimental data, where the unknown expansion coefficients are the fitting
parameters. Such a parametrization will have the correct analytic structure.
After fitting the data given at real energies, one can use the semi-analytic
Jost function to search for resonances in the nearby domain of the Riemann
surface. The efficiency and accuracy of the suggested expansion is demonstrated
by an example of a two-dimensional model potential.

\appendix
\begin{center}
   \bf APPENDICES
\end{center}
\section{Two-dimensional partial-wave decomposition}
\label{app.decomposition}
The partial-wave decomposition of the wave function, scattering amplitude, and
cross section for a particle moving on a plane, is done using the
cylindrical coordinates where the $z$-axis (perpendicular to the plane) is
needed to define the orbital angular momentum. All the steps of such a
decomposition are similar to the three-dimensional case, but the resulting
formulae are not obvious and cannot be easily obtained from the corresponding
$3D$-analysis. The derivations of various formulae of this type are given in
several different papers (see, for example, Refs. \cite{Lapidus, Adhikari}).
Usually these derivations are very concise with many details omitted.
Since such derivations are not present in the standard textbooks on quantum
mechanics, we feel that it is worthwhile to collect everything in one place.
This is why we include this Appendix.

\subsection{Radial Schr\"odinger equation}
\label{app.SchrEq}
\begin{figure}[ht!]
\centerline{\epsfig{file=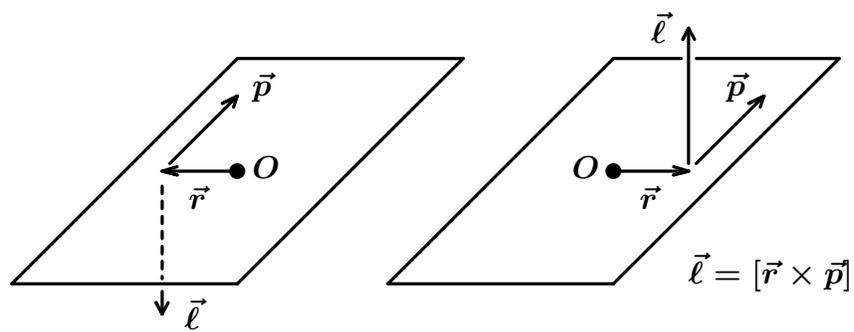}}
\caption{\sf
Two possible directions of the angular momentum for a particle moving on a
plane.
}
\label{fig.vectors}
\end{figure}

Consider a particle of mass $\mu$, moving on a plane and being affected by a
force that is described by a  potential $U(\vec{r})$, which is assumed to be of
a short-range and circularly symmetric,
$$
        U(\vec{r})=U(|\vec{r}|)\ .
$$
To make the derivations simple, we assume that our particle does not have
spin (this restriction can be easily revoked later). In the coordinate
representation, the Hamiltonian $H$ of such a particle is most conveniently
expressed using the polar coordinates,
\begin{eqnarray*}
   H &=& -\frac{\hbar^2}{2\mu}\Delta+U\ ,\\[3mm]
   \Delta &=& \frac{1}{r}\frac{\partial}{\partial r}
   \left(r \frac{\partial}{\partial r}\right) +\frac{1}{\hbar^2r^2}
   \hat{L}^2\ ,
\end{eqnarray*}
where
$$
   \hat{L}^2=\hbar^2\frac{\partial^2}{\partial\varphi^2}
$$
is the two-dimensional operator of the square of the angular
momentum. Its eigenfunctions ${\cal Y}_m(\varphi)$ obeying the equation
$$
   \hat{L}^2{\cal Y}_m(\varphi)=-\hbar^2m^2{\cal Y}_m(\varphi)
$$
and normalized as
\begin{equation}
\label{A.norm}
   \int_0^{2\pi}{\cal Y}_m^*(\varphi){\cal Y}_{m'}(\varphi)d\varphi
   =
   \delta_{mm'}\ ,
\end{equation}
are easy to find,
\begin{equation}
\label{A.Ym}
   {\cal Y}_m(\varphi) = \frac{1}{\sqrt{2\pi}}e^{im\varphi}\ ,
   \qquad
   m=0,\pm1,\pm2,\dots
\end{equation}
From the definition of the Fourier series on the interval $[0,2\pi]$ it follows
that
\begin{equation}
\label{A.sumYdelta}
   \sum_{m=-\infty}^{+\infty}
   {\cal Y}_{m}(\varphi){\cal Y}^*_{m}(\varphi')=\delta(\varphi-\varphi')\ .
\end{equation}
Each value of the angular momentum (except for zero) is represented
twice: with two opposite signs. Classically, these two states correspond to the
motion of the particle at the same distance $r$ from the center and with the
same velocity, but at different sides of the center (see Fig.
\ref{fig.vectors}).\\

The quantum number of the angular momentum $\ell=|m|$ is always
non-negative and irrespective of its magnitude (if $\ell\neq 0$) the vector
$\vec{\ell}$ can have two (only two) directions: up or down
(like the spin 1/2). The quantum number $m=\pm\ell$ is its $z$-component. In
principle, we can use the same notation for the eigenfunctions of the
operator $\hat{L^2}$ as in the three-dimensional case, namely, ${\cal
Y}_{\ell m}$ with two subscripts. However, because of the relation
$m=\pm\ell$, the subscript $\ell$ is redundant.\\

This can be formulated in a different way. The functions (\ref{A.Ym}) form a
complete ortho-normal set on the interval $\varphi\in[0,2\pi]$. This means that
any (reasonable) function $f(\varphi)$ defined on this interval, can be written
as their linear combination, and such a combination can be written in the
following two (equivalent) ways
$$
   f(\varphi)=\sum_{m=-\infty}^{+\infty}a_m{\cal Y}_m(\varphi)=
   \sum_{\ell=0}^\infty\sum_{m=\pm\ell}\tilde{a}_{\ell m}{\cal Y}_m(\varphi)\ ,
   \qquad
   a_m=\tilde{a}_{|m|m}\ .
$$
In other words the following summations are equivalent
$$
   \sum_{m=-\infty}^{+\infty}\ \longleftrightarrow
   \ \sum_{\ell=0}^\infty\sum_{m=\pm\ell}\ .
$$
The operator corresponding to the quantum number $m$ is obtained as follows.
The gradient operator in the cylindrical coordinates is
\begin{equation}
\label{A.nabla}
   \vec{\nabla}=\hat{\vec{r}}\frac{\partial}{\partial r}+
   \hat{\vec{\varphi}}\frac{1}{r}\frac{\partial}{\partial \varphi}+
   \hat{\vec{z}}\frac{\partial}{\partial z}\ ,
\end{equation}
where
$$
   \hat{\vec{r}}=[\hat{\vec{\varphi}}\times\hat{\vec{z}}]\ ,
   \quad
   \hat{\vec{\varphi}}=[\hat{\vec{z}}\times\hat{\vec{r}}]\ ,
   \quad
   \hat{\vec{z}}=[\hat{\vec{r}}\times\hat{\vec{\varphi}}]
$$
are the corresponding unit vectors. Then
$$
   [\vec{r}\times\vec{p}]=
   r\hat{\vec{r}}\times\frac{\hbar}{i}\vec{\nabla}=
   \frac{\hbar r}{i}\left(
   \hat{\vec{z}}\frac{1}{r}\frac{\partial}{\partial \varphi}-
   \hat{\vec{\varphi}}\frac{\partial}{\partial z}\right)
$$
and thus
$$
   \ell_z=\frac{\hbar}{i}\frac{\partial}{\partial \varphi}\ .
$$
Apparently, both $\hat{L}^2$ and $\ell_z$ commute with the Hamiltonian. The
quantum numbers $\ell$ and $m$ are therefore conserving. When specifying
$m$, we implicitly specify the quantum number $\ell$ as well. This means
that the quantum state of the particle is determined by two conserving
quantum numbers, namely, the energy $E$ and the $z$-component of the angular
momentum $m$. The corresponding wave function, obeying the Schr\"odinger
equation,
$$
    H\psi_{Em}(\vec{r})=E\psi_{Em}(\vec{r})\ ,
$$
can be factorized in the radial and angular parts
\begin{equation}
\label{psi_r_phi}
    \psi_{Em}(\vec{r})=\frac{u_m(E,r)}{\sqrt{r}}{\cal Y}_m(\varphi)\ ,
\end{equation}
where $\sqrt{r}$ in the denominator is introduced to obtain the radial
equation without the first derivative. Substituting this factorized form
into the Schr\"odinger equation, we obtain
\begin{equation}
\label{Schradial}
   \left[
   \frac{d^2}{dr^2}+k^2-\frac{m^2-1/4}{r^2}-V(r)\right]
   u_m(E,r)=0\ ,
\end{equation}
where $k$ is the wave number (linear momentum) defined by
\begin{equation}
\label{k2E}
    k^2=\frac{2\mu}{\hbar^2}E
\end{equation}
and $V(r)$ is the reduced (in the units of $\mathrm{[length]}^{-2}$) potential
$$
    V(r)=\frac{2\mu}{\hbar^2}U(r)\ .
$$
Noting that Eq. (\ref{Schradial}) is exactly the same for both choices of the
sign for $m$, we conclude that $u_m(E,r)$ actually depends on $\ell$ but not on
$m$. The radial equation can therefore be re-written in the way we used to see
it in the three-dimensional problems
\begin{equation}
\label{A.radialeq}
   \left[
   \frac{d^2}{dr^2}+k^2-\frac{\lambda(\lambda+1)}{r^2}-V(r)\right]
   u_\ell(E,r)=0\ ,
\end{equation}
where we introduced
\begin{equation}
\label{A.lambda.def}
   \lambda=\ell-\frac12
\end{equation}
and did the replacement
$$
   \ell^2-\frac14=
   \left(\ell-\frac12\right)\left(\ell+\frac12\right)=\lambda(\lambda+1)\ .
$$
Formally, Eq. (\ref{A.radialeq}) looks exactly like the radial equation of the
three-dimensional problem. The only difference is that $\lambda$ is not an
integer number
$$
   \lambda= -\frac12,\ \frac12,\ \frac32,\ \frac52,\ \dots
$$
This simple fact makes a huge difference: it changes the
analytic properties of the Jost function and thus the $S$-matrix, because the
Riccat-Neumann function $y_\lambda(kr)$ with a half-integer $\lambda$ has a
logarithmic
branching point on the Riemann surface of the energy \cite{Abramowitz}.

\subsection{Plane-wave and circular waves}
\label{app.planewave}
Consider a two-dimensional plane wave normalized to the $\delta$-function
\begin{equation}
\label{A.pwnormalized}
   \langle\vec{r}|\vec{k}\rangle=
   \displaystyle\frac{e^{i\vec{k}\vec{r}}}{2\pi}\ ,\qquad
   \langle\vec{k}'|\vec{k}\rangle=\delta\left(\vec{k}'-\vec{k}\right)=
   \frac{1}{k}\delta(k'-k)\delta(\varphi'-\varphi)\ ,
\end{equation}
where $\varphi$ is the polar angle of the momentum $\hbar\vec{k}$.
This plane wave can be expanded over the full set $\{{\cal Y}\}$
of the angular functions (\ref{A.Ym}),
\begin{equation}
\label{A.pwexp}
    \frac{e^{i\vec{k}\vec{r}}}{2\pi}=
    \frac{e^{ikr\cos\varphi}}{2\pi}=
    \sum_{\ell m}a_m(kr){\cal Y}_m(\varphi)\ ,
\end{equation}
where the $x$-axis is directed along the coordinate vector $\vec{r}$.
The expansion coefficients
\begin{equation}
\label{A.am_int}
   a_m(kr)=\frac{1}{(2\pi)^{3/2}}\int_0^{2\pi}
   e^{i(kr\cos\varphi-m\varphi)}d\varphi
\end{equation}
can be found using the integral representation of the Bessel function
\cite{Abramowitz}
\begin{eqnarray}
\nonumber
   J_m(z) &=&
   \frac{1}{\pi i^m}\int_0^\pi e^{iz\cos\varphi}\cos(m\varphi)d\varphi
   =
   \frac{1}{2\pi i^m}\int_0^\pi e^{iz\cos\varphi}\left(
   e^{im\varphi}+e^{-im\varphi}\right)d\varphi\\[3mm]
   &=&
\label{A.Jm_int}
   \frac{1}{2\pi i^m}\int_{-\pi}^\pi
   e^{i(z\cos\varphi-m\varphi)}d\varphi
   =\frac{i^m}{2\pi}\int_0^{2\pi}e^{i(-z\cos\varphi-m\varphi)}d\varphi\ .
\end{eqnarray}
Comparing Eq. (\ref{A.am_int}) with (\ref{A.Jm_int}) and using the symmetry
property of the Bessel function $J_m(-z)=(-1)^mJ_m(z)$, we see that
\begin{equation}
\label{A.am_Jm}
   a_m(kr)=\frac{i^m}{\sqrt{2\pi}}J_m(kr)
\end{equation}
and thus
\begin{equation}
\label{A.pw_bess}
   \frac{e^{i\vec{k}\vec{r}}}{2\pi}=
   \frac{1}{2\pi}\sum_{\ell m}i^me^{im\varphi}J_m(kr)=
   \frac{1}{2\pi}\sum_{-\infty}^{+\infty}i^me^{im\varphi}J_m(kr)=
   \frac{1}{\sqrt{2\pi}}\sum_{-\infty}^{+\infty}i^mJ_m(kr){\cal Y}_m(\varphi)\ .
\end{equation}
Using another symmetry property, $J_{-m}(z)=(-1)^mJ_m(z)$, we see that the
product $i^mJ_m(kr)$ does not depend on the sign of $m$ and thus this expansion
can be re-written as
\begin{eqnarray}
\nonumber
   \frac{e^{i\vec{k}\vec{r}}}{2\pi}
   &=&
   \frac{1}{2\pi}\left[
   J_0(kr)+\sum_{\ell=1}^\infty i^\ell\left(e^{i\ell\varphi}+
   e^{-i\ell\varphi}\right)J_\ell(kr)\right]\\[3mm]
   &=&
\label{A.pw_eps}
   \frac{1}{2\pi}\sum_{\ell=0}^\infty \epsilon_\ell
   i^\ell\cos(\ell\varphi)J_\ell(kr)\ ,
\end{eqnarray}
where $\epsilon_\ell$ is the "multiplicity" of an $\ell$-state, i.e. is the
analog of the factor $(2\ell+1)$ of the 3D-case,
\begin{equation}
\label{A.eps_definition}
   \epsilon_\ell=\left\{
   \begin{array}{rcl}
   1 &,& \ell=0\ ,\\
   2 &,& \ell>0\ .
   \end{array}
   \right.
\end{equation}
Expressing the Bessel function via the Riccati-Hankel functions,
$$
   J_\ell(z)=\sqrt{\frac{1}{2\pi z}}\left[
   h_{\ell-1/2}^{(-)}(z)+h_{\ell-1/2}^{(+)}(z)\right]\ ,
$$
we obtain the following decomposition of the $2D$ plane wave in the
incoming $(-)$ and outgoing $(+)$ circular waves
\begin{equation}
\label{A.pw.in.circular}
   \frac{e^{i\vec{k}\vec{r}}}{2\pi} = \frac{1}{2\pi\sqrt{kr}}
   \sum_{\ell=0}^\infty
   \sum_{m=\pm\ell} i^\ell\left[
   h_{\lambda}^{(-)}(kr)+h_{\lambda}^{(+)}(kr)\right]{\cal Y}_m(\varphi)\ ,
\end{equation}
where $\lambda$ is defined by Eq. (\ref{A.lambda.def}).\\

In the above, we assumed that vector $\vec{r}$ was directed along the $x$-axis.
If this is not the case, then the dot-product
$$
   \vec{k}\vec{r}=kr(\cos\varphi_k\cos\varphi_r+\sin\varphi_k\sin\varphi_r)=
   kr\cos(\varphi_r-\varphi_k)
$$
depends on the two polar angles. In this general case the plane wave is
expanded over two sets of functions $\{{\cal Y}(\varphi_k)\}$ and
$\{{\cal Y}(\varphi_r)\}$ depending on the angles of the momentum and
co-ordinate vectors. In a similar way as we did it above, it is not difficult
to show that
\begin{eqnarray}
\nonumber
   \frac{e^{i\vec{k}\vec{r}}}{2\pi} &=&
   \sum_{-\infty}^{+\infty}i^mJ_m(kr)
   {\cal Y}^*_m(\varphi_r){\cal Y}_m(\varphi_k)=
   \sum_{\ell=0}^\infty
    i^\ell J_\ell(kr)\!\!\sum_{m=\pm\ell}
   {\cal Y}^*_m(\varphi_r){\cal Y}_m(\varphi_k) \\[3mm]
\nonumber
   &=&
   \frac{1}{2\pi}\sum_{\ell=0}^\infty\epsilon_\ell i^\ell
   \cos\left[\ell(\varphi_r-\varphi_k)\right]J_\ell(kr)\\[3mm]
\nonumber
   &=&
   \frac{1}{\sqrt{2\pi kr}}\sum_{\ell m}i^\ell
   \left[h_\lambda^{(-)}(kr)+h_\lambda^{(+)}(kr)\right]
   {\cal Y}^*_m(\varphi_r){\cal Y}_m(\varphi_k)\\[3mm]
\label{A.pw_bessYY}
   &=&
   \frac{1}{\sqrt{2\pi kr}}\sum_{\ell m}i^\ell
   u_\ell^{(0)}(E,r)
   {\cal Y}^*_m(\varphi_r){\cal Y}_m(\varphi_k)\ ,
\end{eqnarray}
where
\begin{equation}
\label{A.u_zero}
   u_\ell^{(0)}(E,r)=h_\lambda^{(-)}(kr)+h_\lambda^{(+)}(kr)=2j_\lambda(kr)
\end{equation}
is a regular solution of the radial Schr\"odinger equation (\ref{A.radialeq})
for the case $V(r)\equiv 0$.
These partial-wave decompositions can be conveniently written in the following
symbolic form
\begin{equation}
\label{A.k_r_bra}
   |\vec{k}\rangle = \sum_{\ell m}|k\ell m\rangle
   {\cal Y}_m(\varphi_k)\ ,\qquad
   |\vec{r}\rangle = \sum_{\ell m}|r\ell m\rangle
   {\cal Y}_m(\varphi_r)\ ,
\end{equation}
\begin{equation}
\label{A.rlmklm}
   \langle r\ell m|k\ell'm'\rangle =
   \delta_{\ell\ell'}\delta_{mm'}i^\ell J_\ell(kr)
   =
   \delta_{\ell\ell'}\delta_{mm'}i^\ell\sqrt{\frac{2}{\pi kr}}
   j_\lambda(kr)\ ,
\end{equation}
\begin{equation}
\label{A.rk_kr_lm}
   \langle\vec{r}|k\ell m\rangle =
   i^\ell\sqrt{\frac{2}{\pi kr}}j_\lambda(kr){\cal Y}^*_m(\varphi_r)\ ,
   \qquad
   \langle\vec{k}|r\ell m\rangle =
   (-i)^\ell\sqrt{\frac{2}{\pi kr}}j_\lambda(kr)
   {\cal Y}^*_m(\varphi_k)\ ,
\end{equation}
\begin{equation}
\label{A.klmklm_rlmrlm}
   \langle k\ell m|k'\ell'm'\rangle =
   \frac{1}{k}\delta(k-k')\delta_{\ell\ell'}\delta_{mm'}\ ,
   \qquad
   \langle r\ell m|r'\ell'm'\rangle =
   \frac{1}{r}\delta(r-r')\delta_{\ell\ell'}\delta_{mm'}\ ,
\end{equation}
\begin{equation}
\label{A.closure}
   \int_0^\infty\sum_{\ell m}|k\ell m\rangle\langle k\ell m|k\,dk=1\ ,
   \qquad
   \int_0^\infty\sum_{\ell m}|r\ell m\rangle\langle r\ell m|r\,dr=1\ .
\end{equation}

\subsection{Scattering wave function}
\label{app.scatt.wf}
The plane wave (\ref{A.pw_bessYY}) is a scattering wave function
$\psi_{\vec{k}}(\vec{r})$ for the particular case of $V(r)\equiv 0$. Apparently,
the structure of its partial-wave decomposition should be the same for all
potentials
$$
   \psi_{\vec{k}}(\vec{r})=
   \frac{N}{\sqrt{2\pi kr}}\sum_{\ell m}i^\ell
   u_\ell(E,r)
   {\cal Y}^*_m(\varphi_r){\cal Y}_m(\varphi_k)\ ,
$$
where the factor $N$ is determined by the choice of the potential and the
collision energy (for the free motion, $N=1$ at all energies). The purpose of
this factor is to always have exactly the same normalization, namely,
\begin{equation}
\label{A.delta_normalization}
   \langle \psi_{\vec{k}}|\psi_{\vec{k}'}\rangle=
   \delta\left(\vec{k}-\vec{k}'\right)\ .
\end{equation}
An appropriate value for $N$ can be found as follows. The Riccati-Hankel
functions $h_\lambda^{(\pm)}(kr)$ are two linearly
independent solutions of the radial Schr\"odinger equation (\ref{A.radialeq})
without the potential term. This means that for a short-range potential, its
solution asymptotically behaves as a linear combination of the Riccati-Hankel
functions,
\begin{equation}
\label{A.uASS}
   u_\ell(E,r)
   \ \mathop{\longrightarrow}\limits_{r\to\infty}
   \ f_\ell^{(\mathrm{in})}(E)h_{\lambda}^{(-)}(kr)+
     f_\ell^{(\mathrm{out})}(E)h_{\lambda}^{(+)}(kr)\ ,
\end{equation}
where the combination coefficients depend on the energy and are called the Jost
functions. In fact, they are the amplitudes of the incoming $(-)$ and outgoing
$(+)$ circular waves. On the other hand, at large distances the wave function
$\psi_{\vec{k}}(\vec{r})$ consists of the two parts: the initial (incident)
wave $\psi_{\vec{k}}^{(0)}(\vec{r})$ and a scattered circular wave
that goes in all directions with certain amplitude $F$,
$$
   \psi_{\vec{k}}(\vec{r})\ \mathop{\longrightarrow}\limits_{r\to\infty}
   \ \psi_{\vec{k}}^{(0)}(\vec{r})+F(E,\varphi_k,\varphi_r)
   \frac{e^{ikr}}{\sqrt{r}}\ .
$$
If $\psi_{\vec{k}}(\vec{r})$ is properly normalized then
$\psi_{\vec{k}}^{(0)}(\vec{r})\equiv {e^{i\vec{k}\vec{r}}}/{(2\pi)}$. This
means that the radial wave function at large distances should also be split in
two parts one of which coincides with the function (\ref{A.u_zero}). In doing
such a splitting of the function (\ref{A.uASS}), we obtain
$$
   u_\ell(E,r)
   \ \mathop{\longrightarrow}\limits_{r\to\infty}
   \ f_\ell^{(\mathrm{in})}\left[h_{\lambda}^{(-)}+h_{\lambda}^{(+)}+
   \left(\frac{f_\ell^{(\mathrm{out})}}{f_\ell^{(\mathrm{in})}}-1\right)
   h_{\lambda}^{(+)}\right]
$$
and see that $N=1/f_\ell^{(\mathrm{in})}$, i.e.
\begin{equation}
\label{A.ScWF_normalized}
   \psi_{\vec{k}}(\vec{r})=
   \frac{1}{\sqrt{2\pi kr}f_\ell^{(\mathrm{in})}(E)}\sum_{\ell m}i^\ell
   u_\ell(E,r)
   {\cal Y}^*_m(\varphi_r){\cal Y}_m(\varphi_k)\ .
\end{equation}

\subsection{Cross section}
\label{app.scatt.cross.section}
Defining the partial-wave $S$-matrix and the amplitude,
$$
   s_\ell(E)=\frac{f_\ell^{(\mathrm{out})}(E)}{f_\ell^{(\mathrm{in})}(E)}\ ,
   \qquad
   f_\ell(E)=\frac{s_\ell(E)-1}{\sqrt{2\pi ik}}\ ,
$$
and using
$$
   h_\lambda^{(+)}(kr)\ \mathop{\longrightarrow}\limits_{r\to\infty}
   \ -i\exp\left[i(kr-\lambda\pi/2)\right]\ ,
$$
as well as the fact that
\begin{eqnarray*}
   \sum_{m}
   {\cal Y}^*_m(\varphi_r){\cal Y}_m(\varphi_k)
   &=&
   \left\{
   \begin{array}{lcl}
   \displaystyle
   \frac{1}{2\pi} &,& \ell=0\\[3mm]
   \displaystyle
   \frac{1}{2\pi}\left(e^{-i\ell\varphi_r}e^{i\ell\varphi_k}+
   e^{i\ell\varphi_r}e^{-i\ell\varphi_k}\right) &,& \ell>0
   \end{array}\right.\\[3mm]
   &=&
   \frac{\epsilon_\ell}{2\pi}\cos\left[\ell(\varphi_k-\varphi_r)\right]\ ,
\end{eqnarray*}
we can write the asymptotic behaviour of the scattering wave function as
\begin{equation}
\label{A.WF_ass_vect}
   \psi_{\vec{k}}(\vec{r})
   \ \mathop{\longrightarrow}\limits_{r\to\infty}
   \ \frac{1}{2\pi}\left[e^{i\vec{k}\vec{r}}+
   {\cal F}(E,\varphi)\frac{e^{ikr}}{\sqrt{r}}\right]\ ,
\end{equation}
where $\varphi=\varphi_k-\varphi_r$ is the scattering angle and the total
scattering amplitude has the following partial-wave expansion
\begin{equation}
\label{A.F_pw_expansion}
   {\cal F}(E,\varphi)=
   \sum_{\ell=0}^\infty
   \epsilon_\ell f_\ell(E)\cos(\ell\varphi)\ .
\end{equation}
The cross section for a two-dimensional scattering has the units of length.
The number of particles scattered into the angle spanned by the arc
$r\,d\varphi$, is the product of the flux in that radial direction and the
length of the arc. The corresponding cross section $d\sigma$ is
defined as such a length that after its multiplication by the total incoming
flux, it gives the same number of particles, i.e.
$$
     \left|\vec{j}^{(\mathrm{in})}\right|d\sigma=
     j_r^{(\mathrm{out})}(\varphi)rd\varphi\ .
$$
Using standard definition for the particle flux
$\vec{j}=\hbar/(2i\mu)(\psi^*\vec{\nabla}\psi-\psi\vec{\nabla}\psi^*)$ with the
operator $\vec{\nabla}$ given by Eq. (\ref{A.nabla}), it is not difficult to
find that the incoming [corresponding to the first term of the
wave function (\ref{A.WF_ass_vect})] and outgoing (obtained from the second
term of the same wave function) fluxes are
$$
    \vec{j}^{(\mathrm{in})}=\frac{\hbar\vec{k}}{(2\pi)^2\mu}\ ,
    \qquad
    j_r^{(\mathrm{out})}(\varphi)=\frac{\hbar k\left|{\cal F}(E,\varphi)
    \right|^2}{(2\pi)^2\mu r}
$$
and thus the differential cross section is
$$
   \frac{d\sigma}{d\varphi}=\left|{\cal F}(E,\varphi)\right|^2\ .
$$
Using the integral
$$
   \int_0^{2\pi}\cos(\ell\varphi)\cos(\ell'\varphi)d\varphi=
   \left\{\begin{array}{lcl}
   0 &,& \ell\neq\ell'\\
   2\pi &,& \ell=\ell'=0\\
   \pi &,& \ell=\ell'\neq 0
   \end{array}
   \right\}=\frac{2\pi}{\epsilon_\ell}\delta_{\ell\ell'}
$$
the total cross section can be written as follows
\begin{eqnarray}
\nonumber
   \sigma &=& \int_0^{2\pi}\left|{\cal F}(E,\varphi)\right|^2d\varphi
   =
\label{A.partial_cross_section}
   \sum_\ell \sigma_\ell\ ,\\[3mm]
   \sigma_\ell &=&
   2\pi\epsilon_\ell\left|f_\ell(E)\right|^2
   =\frac{\epsilon_\ell}{k}\left|s_\ell(E)-1\right|^2\ ,
\end{eqnarray}
where $\sigma_\ell$ is the partial-wave cross section.

\section{Expansion coefficients for the holomorphic parts of the Riccati-Bessel
and Riccati-Neumann functions}
\label{app.coefficients}
The Riccati-Bessel and Riccati-Neumann functions $j_\lambda(kr)$ and $y_\lambda(kr)$
can be written in the following factorized form
\begin{eqnarray}
\label{A:jfact}
    j_\lambda(kr) &=&
    k^{\lambda+1}\tilde{j}_\lambda(E,r)\ ,\\[3mm]
\label{A:yfact}
    y_\lambda(kr) &=&
    k^{-\lambda}\tilde{y}_\lambda(E,r)+
    k^{\lambda+1}h(k)\tilde{j}_\lambda(E,r)\ ,
\end{eqnarray}
where the "tilded" functions are holomorphic with respect to the energy variable $E$.
This means that we can expand them in the Taylor series,
\begin{eqnarray}
\label{A:jtildeE0}
    \tilde{j}_\lambda(E,r) &=&
    \sum_{n=0}^\infty
    s_n^{(\lambda)}(E_0,r)\left(E-E_0\right)^n\ ,\\[3mm]
\label{A:ytildeE0}
    \tilde{y}_\lambda(E,r) &=&
    \sum_{n=0}^\infty
    c_n^{(\lambda)}(E_0,r)\left(E-E_0\right)^n\ ,
\end{eqnarray}
near an arbitrary point $E_0$. The expansion coefficients,
\begin{eqnarray}
\label{A:sn}
    s_n^{(\lambda)}(E_0,r) &=&
    \left.
    \frac{1}{n!}\frac{\partial^n}{\partial E^n}
    \tilde{j}_\lambda(E,r)\right|_{E=E_0}\ ,\\[3mm]
\label{A:cn}
    c_n^{(\lambda)}(E_0,r) &=&
    \left.
    \frac{1}{n!}\frac{\partial^n}{\partial E^n}
    \tilde{y}_\lambda(E,r)\right|_{E=E_0}\ ,
\end{eqnarray}
are expressed via the corresponding derivatives. In order to find them, we notice that
\begin{equation}
\label{A:dEdk}
    E=\frac{\hbar^2k^2}{2\mu}
    \quad \Longrightarrow
    \quad \frac{\partial}{\partial E}=
    \frac{\mu}{\hbar^2k}\frac{\partial}{\partial k}
\end{equation}
and also make use of the relations (which follow from Eq.(9.1.30) of the handbook by
M. Abramowitz and I. A. Stegun)
\begin{eqnarray}
\label{A:relation1}
    \frac{d}{dz}\left[\frac{{\cal J}_\lambda(z)}{z^{\lambda+1}}\right]
    &=&
    -\frac{{\cal J}_{\lambda+1}(z)}{z^{\lambda+1}}\ ,\\[3mm]
\label{A:relation2}
    \frac{d}{dz}\left[{z^{\lambda}{\cal J}_\lambda(z)}\right]
    &=&
    z^{\lambda}{\cal J}_{\lambda-1}(z)\ ,
\end{eqnarray}
where ${\cal J}_\lambda(z)$ stands for either $j_\lambda(z)$ or $y_\lambda(z)$. Therefore
\begin{eqnarray*}
   \frac{\partial}{\partial E}\tilde{j}_\lambda(E,r) &=&
   \frac{\mu}{\hbar^2k}\cdot\frac{\partial}{\partial k}\left[
   \frac{j_\lambda(kr)}{k^{\lambda+1}}\right]=
   \frac{\mu r^{\lambda+2}}{\hbar^2k}\cdot
   \frac{\partial}{\partial (kr)}\left[
   \frac{j_\lambda(kr)}{(kr)^{\lambda+1}}\right]=\\[3mm]
   &=&
   -\frac{\mu r^{\lambda+2}}{\hbar^2k}
   \frac{j_{\lambda+1}(kr)}{(kr)^{\lambda+1}}=
   -\frac{\mu r}{\hbar^2} \tilde{j}_{\lambda+1}(E,r)
\end{eqnarray*}
and thus
\begin{equation}
\label{A:djdEn}
   \frac{\partial^n}{\partial E^n}\tilde{j}_\lambda(E,r)
   =
   \left(-\frac{\mu r}{\hbar^2}\right)^n\tilde{j}_{\lambda+n}(E,r)\ ,\\[3mm]
\end{equation}
\begin{equation}
\label{A:snlambda}
   s_n^{(\lambda)}(E_0,r) =
   \frac{1}{n!}\left(-\frac{\mu r}{\hbar^2}\right)^n
   \left[\frac{j_{\lambda+n}(kr)}{k^{\lambda+n+1}}\right]_{E=E_0}
   =
   \frac{1}{n!}\left(-\frac{\mu r}{\hbar^2}\right)^n
   \sqrt{\frac{\pi r}{2}}
   \left[\frac{J_{\ell+n}(kr)}{k^{\ell+n}}\right]_{E=E_0}\ .
\end{equation}
As it should be (since $\tilde{j}_\lambda(E,r)$ is single-valued), the expansion
coefficients $s_n^{(\lambda)}(E_0,r)$ do not depend on the choice of the sign of the
momentum $k_0=\pm\sqrt{2\mu E_0/\hbar^2}$. Indeed,
\begin{equation}
\label{A:ratio}
   \frac{j_{\lambda+n}(kr)}{k^{\lambda+n+1}}=
   \sqrt{\frac{\pi kr}{2}}\frac{J_{\lambda+n+1/2}(kr)}{k^{\lambda+n+1}}=
   \sqrt{\frac{\pi r}{2}}\frac{J_{\lambda+n+1/2}(kr)}{k^{\lambda+n+1/2}}\ ,
\end{equation}
and since (see Eq.(9.1.35) of M. Abramowitz et al.)
\begin{equation}
\label{A:Jsign}
   J_\nu(ze^{i\pi})=\left(e^{i\pi}\right)^\nu J_\nu(z)\ ,
\end{equation}
the numerator and denominator in Eq. (\ref{A:ratio}) acquire the same phase factor
when $k$ changes its sign.\\

Finding the derivative $\partial^n_E\tilde{y}_\lambda(E,r)$ is a little bit more
complicated. The first derivative can be written as
\begin{eqnarray*}
   \frac{\partial}{\partial E}\tilde{y}_\lambda(E,r)
   &=&
   \frac{\mu r^{1-\lambda}}{\hbar^2k}\left\{
   \frac{\partial}{\partial(kr)}
   \left[(kr)^\lambda y_\lambda(kr)\right]-
   \frac{\partial}{\partial(kr)}
   \left[h(k)(kr)^\lambda j_\lambda(kr)\right]\right\}\ .
\end{eqnarray*}
Using Eq. (\ref{A:relation2}) and explicit form of the function $h(k)$ given by
Eq. (\ref{hnonzero}), we obtain
\begin{eqnarray}
\nonumber
   \frac{\partial}{\partial E}\tilde{y}_\lambda(E,r)
   &=&
   \frac{\mu r^{1-\lambda}}{\hbar^2k}\left[
   (kr)^\lambda y_{\lambda-1}(kr)-
   h(k)(kr)^\lambda j_{\lambda-1}(kr)-
   (kr)^\lambda j_\lambda(kr)\frac{2}{\pi kr}
   \right]\\[3mm]
\nonumber
   &=&
   \frac{\mu r}{\hbar^2}\left[k^{\lambda-1}y_{\lambda-1}(kr)-
   k^{\lambda-1}h(k)j_{\lambda-1}(kr)-
   \frac{2}{\pi r}k^{\lambda-2}j_\lambda(kr)\right]\\[3mm]
\nonumber
   &=&
   \frac{\mu r}{\hbar^2}\tilde{y}_{\lambda-1}(E,r)-
   \frac{2\mu}{\pi\hbar^2}k^{\lambda-2}j_\lambda(kr)\\[3mm]
\label{A:DyDE}
   &=&
   \frac{\mu r}{\hbar^2}\tilde{y}_{\lambda-1}(E,r)-
   \frac{2\mu}{\pi\hbar^2}f_{\lambda 1}(kr)\ ,
\end{eqnarray}
where we introduced an auxiliary function
\begin{equation}
\label{A:flambdan}
   f_{\lambda n}(k,r)=k^{\lambda-2n}j_\lambda(kr)
   =k^{\ell-2n}\sqrt{\frac{\pi r}{2}}J_\ell(kr)\ ,
\end{equation}
whose derivatives can be found using the following recurrence relation
\begin{eqnarray*}
   \frac{\partial}{\partial E}\left[
   k^{\lambda-2n}j_\lambda(kr)\right] &=&
   \frac{\mu}{\hbar^2k}r^{2n-\lambda+1}
   \frac{\partial}{\partial(kr)}\left[
   \frac{1}{(kr)^{2n}}(kr)^\lambda j_\lambda(kr)\right]\\[3mm]
   &=&
   \frac{\mu}{\hbar^2k}r^{2n-\lambda+1}\left[
   -\frac{2n}{(kr)^{2n+1}}(kr)^\lambda j_\lambda(kr)+
   \frac{1}{(kr)^{2n}}(kr)^\lambda j_{\lambda-1}(kr)\right]\\[3mm]
   &=&
   -\frac{2n\mu}{\hbar^2}k^{\lambda-2(n+1)}j_\lambda(kr)+
   \frac{\mu r}{\hbar^2}k^{\lambda-1-2n}j_{\lambda-1}(kr)\ ,
\end{eqnarray*}
i.e.
\begin{equation}
\label{A:frecurrence}
   \frac{\partial}{\partial E}f_{\lambda n}=
   -\frac{2n\mu}{\hbar^2}f_{\lambda,n+1}+\frac{\mu r}{\hbar^2}
   f_{\lambda-1,n}\ .
\end{equation}
Repeatedly using the relations (\ref{A:DyDE}) and (\ref{A:frecurrence}), we can calculate
any number of the derivatives $\partial^n_E\tilde{y}_\lambda(E,r)$ needed for finding the
expansion coefficients (\ref{A:cn}).


%
%
\begin{table}
\boldmath
\begin{center}
\begin{tabular}{|r|r|}
\hline
\multicolumn{1}{|c|}{$E_r$} & \multicolumn{1}{|c|}{$\Gamma$} \\
\hline
\multicolumn{1}{|r|}{$-32.4850428093$} & \multicolumn{1}{|c|}{$0$}\\
\hline
\multicolumn{1}{|r|}{$-16.2643650096$} & \multicolumn{1}{|c|}{$0$}\\
\hline
\multicolumn{1}{|r|}{$ -6.2711504590$} & \multicolumn{1}{|c|}{$0$}\\
\hline
\multicolumn{1}{|r|}{$0.5036180960$} & \multicolumn{1}{|c|}{$2\times10^{-15}$}\\
\hline
$4.9422440057$ & $0.0000588188$\\
\hline
$7.1050168573$ & $0.5710776714$\\
\hline
$7.9987409699$ & $3.6684977768$\\
\hline
$8.5025637363$ & $7.7605743107$\\
\hline
$8.5937554145$ & $12.3052581635$\\
\hline
$8.3193121385$ & $17.0922638769$\\
\hline
$7.6952969586$ & $21.9663916836$\\
\hline
$6.7436612278$ & $26.9367555304$\\
\hline
$5.5244040747$ & $31.8688591621$\\
\hline
$4.0100103640$ & $36.8118853195$\\
\hline
$2.2603614329$ & $41.6490284540$\\
\hline
\end{tabular}
\end{center}
\caption{\sf
Spectral points $E=E_r-i\Gamma/2$ (in the units $10.96\,\mathrm{meV}$) generated
by the potential (\ref{Example.V}). Their distribution on the complex energy
surface is shown in Fig.~\protect\ref{fig.spectrum}.
}
\label{table.spectrum}
\end{table}

%
%
\end{document}